\documentclass[submission,copyright,creativecommons]{eptcs}
\providecommand{\event}{ Hybrid Systems and Biology, 2013} 
\usepackage{breakurl}             
\usepackage{stmaryrd}
\usepackage{amsmath}
\usepackage{amssymb}
\usepackage{latexsym}
\usepackage{amsthm}
\usepackage{mathrsfs}
\usepackage{fancyhdr}
\usepackage{caption}
\usepackage{amsfonts}
\usepackage{amssymb}
\usepackage{graphicx}
\usepackage{multirow}
\usepackage{cite}
\usepackage{latexsym , bm}
\usepackage{chemarr}
\usepackage{textgreek}
\makeatletter
\def\ExtendSymbol#1#2#3#4#5{\ext@arrow 0099{\arrowfill@#1#2#3}{#4}{#5}}
\def\RightExtendSymbol#1#2#3#4#5{\ext@arrow 0359{\arrowfill@#1#2#3}{#4}{#5}}
\def\LeftExtendSymbol#1#2#3#4#5{\ext@arrow 6095{\arrowfill@#1#2#3}{#4}{#5}}
\makeatother

\newcommand\scalemath[2]{\scalebox{#1}{\mbox{\ensuremath{\displaystyle #2}}}}
\newcommand\vrr{\mathsf{VRR}}
\newcommand\vr{\mathsf{VR}}
\newcommand\rvr{\mathsf{RVR}}
\newcommand\rmono{\mathsf{R}}
\newcommand\rr{\mathsf{RR}}
\newcommand\vc{\mathsf{V}}
\newcommand\x{_\mathrm{x}}
\newcommand{\bX}{\mathbf{X}}
\newcommand{\bA}{\mathbf{A}}

\newcommand\Rmono{\mathrm{[R]}}

\title{The impact of high density receptor clusters on VEGF signaling}
\author{
Ye Chen
\institute{Department of Mathematics\\ West Virginia University}
\email{chenye@math.wvu.edu}
\and
 Christopher Short
\institute{Department of Mathematics\\ West Virginia University}
\email{cshort3@mix.wvu.edu}
\and
 \'Ad\'am M. Hal\'asz
\institute{Department of Mathematics\\ West Virginia University}
\email{halasz@math.wvu.edu}
\and
Jeremy S. Edwards
\institute{Department of Molecular Genetics and Microbiology\\ University of New Mexico}
\email{jsedwards@salud.unm.edu}
\thanks{This work was supported by NIH grants R01 GM104973 (to JSE and \'AMH)  and K25 CA131558 (\'AMH).}
}
\def\titlerunning{VEGF Signaling with One Receptor Cluster}
\def\authorrunning{Y. Chen, C. Short, \'A. M. Hal\'asz, J. S. Edwards}

\begin{document}
\maketitle

\begin{abstract}
Vascular endothelial growth factor (VEGF) signaling is involved in the process of blood vessel development and maintenance. Signaling is initiated by binding of the bivalent VEGF ligand to the membrane-bound receptors (VEGFR), which in turn stimulates receptor dimerization.  Herein, we discuss experimental evidence that VEGF receptors localize in caveloae and other regions of the plasma membrane, and for other receptors, it has been shown that receptor clustering has an impact on dimerization and thus also on signaling. Overall, receptor clustering is part of a complex ecosystem of interactions and how receptor clustering impacts dimerization is not well understood. To address these questions, we have formulated the simplest possible model. We have postulated the existence of a single high affinity region in the cell membrane, which acts as a transient trap for receptors.  We have defined an ODE model by introducing high- and low-density receptor variables and introduce the corresponding reactions from a realistic model of VEGF signal initiation. Finally, we use the model to investigate the relation between the degree of VEGFR concentration, ligand availability, and signaling. In conclusion, our simulation results provide a deeper understanding of the role of receptor clustering in cell signaling.
\end{abstract}

\section{Introduction}

The topic of the spatial organization of the cell membrane and its impact on receptor clustering and signal initiation are part of a complex and very active field,  illustrating the challenges faced by quantitative systems biology.
There are ultimately {\em two different levels} of spatial and mathematical detail involved. Signaling in response to the presence of VEGF occurs on the level of {\em the entire cell}. Factors that enhance or inhibit signaling are of crucial importance in the quest to understand and control the progression of various types of cancer.
At the other end of the spectrum, the detailed topography of the cell membrane, the mobility and binding characteristics of individual receptors,  occur at scales of a few {\em  tens of nanometers}, literally {\em thousands of times smaller} than the size of the cell. These aspects are actively investigated by various microscopy modalities, which provide a wealth of extremely detailed data.

One of the tasks of meaningful modeling is to bridge the gap between these scales, and to identify rational approaches to abstractions and approximations that can connect data and insights from different scales. Hybrid systems result naturally when continuous degrees of freedom are abstracted into discrete states or regimes.
Here we discuss a continuous model that results as the ultimate abstraction of the complex biological system mentioned above. Our starting point is the microscopic observation that receptors tend to concentrate in small patches. The distribution, size, and physical characteristics of these patches can be inferred from microscopic observations.
In other work \cite{HsiehYRSVSWE2008,MingYREW2010}, we performed detailed, spatial simulations of receptors in a network of high- and low- density membrane patches. These models are naturally abstracted to a network of patches that act as well-mixed, communicating containers. The final abstraction is one where all high density patches are treated as a single, well mixed compartment, in contact with another one, that represents the rest of the membrane.

In this paper we focus on the final, "top level" abstraction, which becomes quite complex when one combines it with a realistic, kinetic model of signal initiation. We use a recently developed approach \cite{HalaszLMRE2013} to identify and investigate the steady states of the model, and discuss the implications of high density patches on the phenomenology of signaling. 

This rest of this paper is organized as follows. We first provide some background on the role of VEGF, its signaling mechanism, and the potential modulation of VEGF signaling by the spatial structure of the cell membrane. The following subsection is devoted to the phenomenology of receptor clustering and the available experimental data. We conclude the introduction by sketching the sequence of abstractions and approximations required to extract high, cell level behaviors from the detailed microscopic observations. Sec. 2 is devoted to the definition of the model, and to the derivation of analytical expressions for the steady states, that require solving a one dimensional algebraic equation. Section 3 discusses results obtained by numerically solving the steady state expressions.

\subsubsection*{Background}
Angiogenesis, the growth of new blood vessels from preexisting vessels, is switched on or off by the dynamic balance among numerous angiogenic stimulators and inhibitors (the 'angiogenesis switch' hypothesis) \cite{HanahanF1996,BirkBMC2010}. Among the various growth factors, vascular endothelial growth factor (VEGF) and its receptors (VEGFR) have received much attention, because of their fundamental role in tumorigenesis and other pathologies \cite{BirkBMC2010,Karamysheva2008,OlssonKC2006}. Initially identified as a vascular permeability factor that increased leakiness of blood vessels \cite{SengerGDPHD1983}, the role of VEGF in regulating angiogenesis was discovered later \cite{FerraraHGN2004,PlouetSG1989}.

\begin{figure}
\captionsetup{font=scriptsize}
\begin{center}
\includegraphics[scale=0.5]{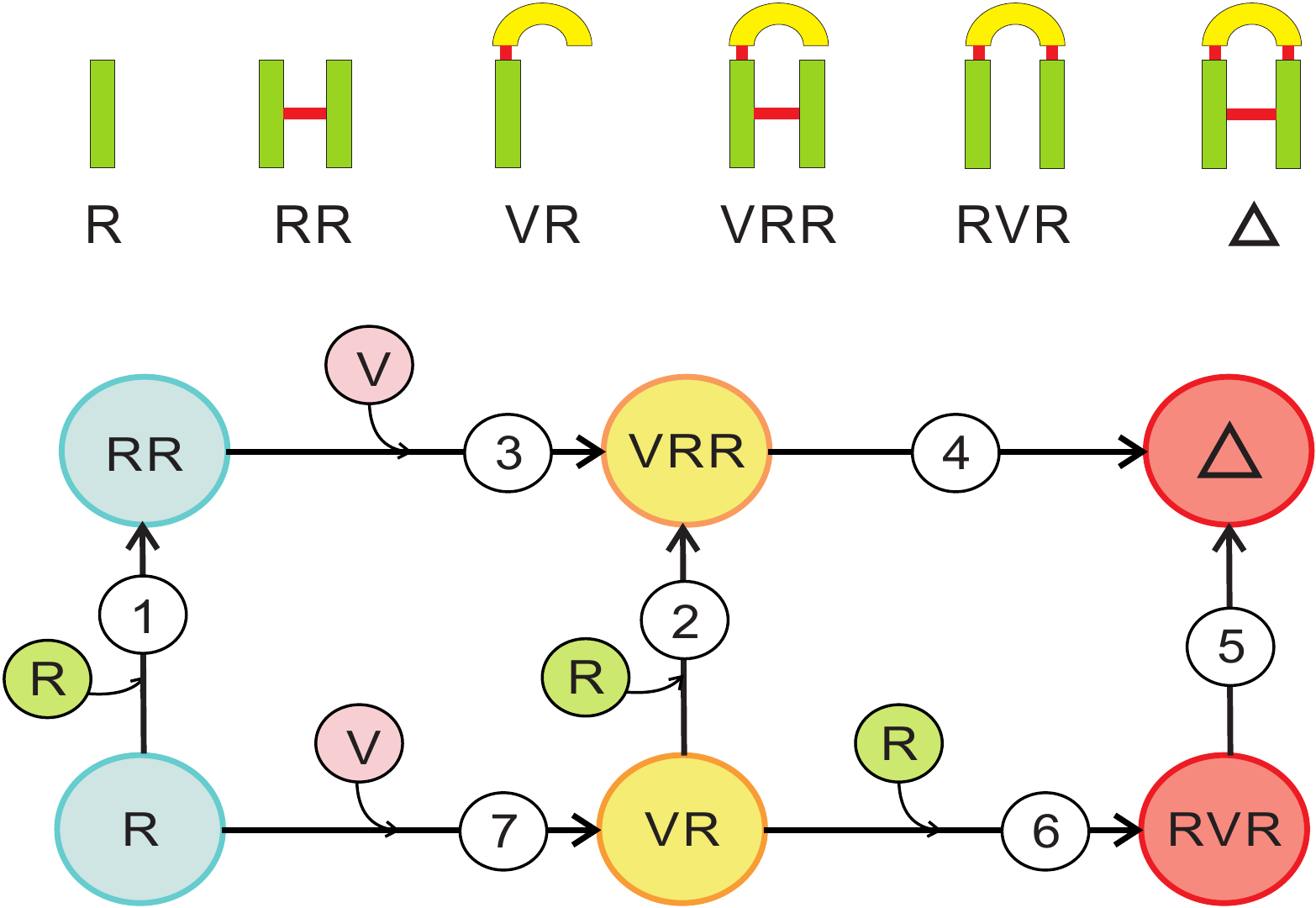}
\end{center}
\caption{The dimerization and ligand binding reactions form a network of 7 reactions in the VEGF signal initiation model of \cite{MacGabhannP2007}.
Receptors ($R$) may bind one of the two poles of a VEGF ligand ($V$), and may form a direct bond with another receptor.
 In the ligand-induced dimerization (LID) sequence, receptors can not form a direct bond outside a pre-existing complex; signal initiation progresses
through reactions ($\mathrm{7,6,5}$).
 In the dynamic pre-dimerization (DPD) sequence, receptors may dimerize before binding ligand (adding reactions $1$ and $2$).
}
\label{7react}
\end{figure}

Signaling by VEGFR is initiated by binding of the ligand dimer to the extracellular domain  of the receptor, which stimulates receptor homo- and hetero-dimerization \cite{LohelaBTA2009,StuttfildB2009,Roskoski2008}. Receptor dimerization is followed by protein kinase activation, trans-autophosphorylation, recruitment of signaling molecules, and activation of distinct pathways. Due to its bivalence, VEGF binding may precede and induce the dimerization of its receptors, by the binding of a second receptor to the free binding site of the ligand (see Figure \ref{7react} for the explicit process).
Ligand-induced or -enhanced receptor dimerization is a feature present in several other receptor-ligand families including EGF and immune receptors.

Mathematical models of VEGF binding \cite{MacGabhannP2007} generally represent the cell membrane as a single, homogeneous entity, equivalent to a "well-mixed compartment'' whose state is sufficiently characterized by a single concentration value for each of the substances of interest. This is justified if there are no significant inhomogeneities and all molecules can diffuse and mix freely over the entire membrane surface, as in the classic Singer-Nicholson fluid mosaic model \cite{SingerN1972}.
 However, our understanding of the cell membrane has evolved significantly since 1972. 
 The current picture \cite{VerbSMNFVMWD2003} is more structured, with microdomains of lipids and proteins \cite{GallegosSKSB2006,SchorederGASM2001,LillemeierPSWD2006}. Modern microscopy techniques \cite{RitchieK2003,WilsonPRLA2007} provide direct evidence of the effect of these structures on membrane receptor localization and movement \cite{ORRHOOW2005,AndrewsLPBW2008,MuraseFUSIM2004}, revealing receptor clusters in static images, and intervals of confinement in small areas separated by jumps or "hops" in single particle tracking.

Spatial organization in the membrane can potentially have a major impact on signaling pathways that rely on interaction between membrane-bound molecules. Receptor dimerization, either through (ligand-dependent or independent) direct receptor-receptor binding, or by crosslinking through ligand[s], requires the collision of two membrane-bound receptors, and is thus influenced by the mobility and possible confinement of receptors. In turn, receptor dimerization is a necessary step in signal initiation, and therefore the mobility and spatial organization of membrane receptors must be part of the quantitative understanding of many cell signaling pathways.

\subsubsection*{The microscopic picture}
VEGF receptors share many properties of other receptor tyrosine kinases. Similarly to EGF receptors, they form ligand-bound dimers in order to activate their intracellular tyrosine kinase domains \cite{LemonS2010}. Experimental and theoretical investigation of EGF binding \cite{MingYREW2010,HsiehYRSVSWE2008,CostaRWVE2009} emphasized the importance of spatial distribution of receptors. Ample experimental evidence indicates that EGF receptors can have a highly inhomogeneous distribution characterized by small areas of high density \cite{YangRYZLSWOW2007}, and exhibit anomalous diffusion \cite{ORRHOOW2005}. There are other examples of receptors that exhibit clustering and anomalous diffusion \cite{AndrewsLPBW2008}.
Receptor accumulation in high density patches has an impact on dimerization and on signaling \cite{CostaRWVE2009,MayawalaVE2005,MayawalaVE22005,MingYREW2010,HsiehYRSVSWE2008}.
\begin{figure}[!ht]
\captionsetup{font=scriptsize}
\centering
\includegraphics[trim=0.0in 0in 0.0in 0.0in, clip, height=0.23\textheight]{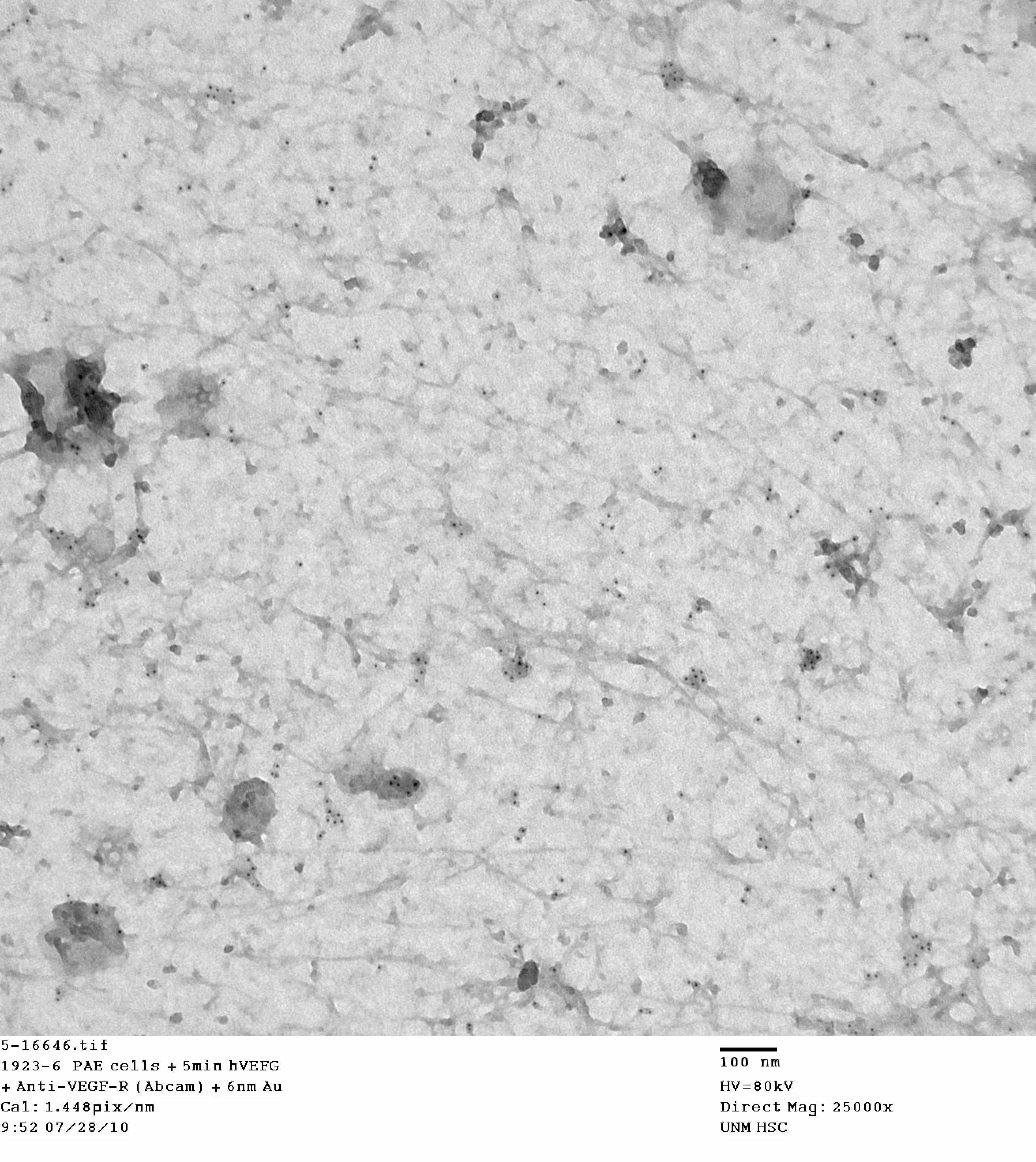}
\includegraphics[trim=0.0in 0in 0.0in 0.0in, clip, height=0.23\textheight]{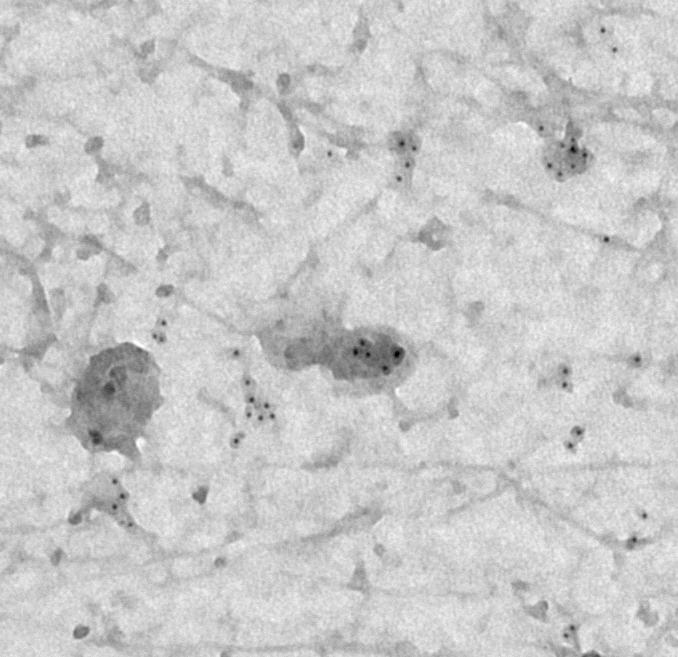}
\includegraphics[trim=0.0in 0in 0.0in 0.0in, clip, height=0.23\textheight]{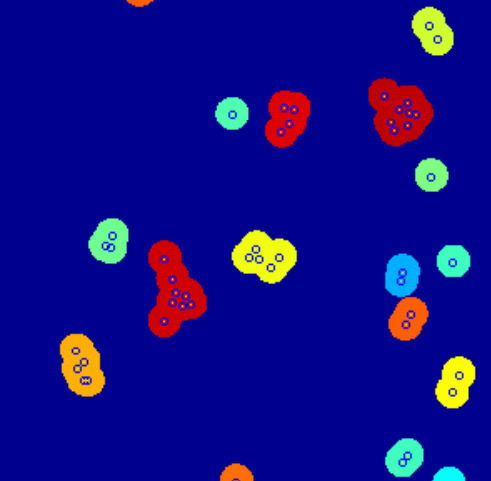}
\caption{ ({\em L}) Transmission electron microscopy (TEM) images of nano-gold labelled VEGF receptors on the membrane of PAE-KDR cells, courtesy of the Wilson lab at UNM.
The second image is a detail of the first one. Gold particles appear as dark spots, whose coordinates are extracted in a semi-automatic procedure. ({\em R}) The result of a distance based cluster identification algorithm applied to the second image.
\label{fig:TEMillustration}
}
\end{figure}

The data analysis pipeline in this case begins with detailed microscopic observations, that provide either static images of a large fraction of the receptors of interest, or, in the case of single particle tracking (SPT), time histories of the positions of a small fraction of the receptors. In the first case (Fig. \ref{fig:TEMillustration}) the imaging modality is transmission electron microscopy (TEM); receptors are tagged with small (6-10 nanometer) sized gold particles, and one image covers a few $\mathrm{\mu m^2}$
capturing up to a few hundred receptors. In the second case (SPT), tracking is typically performed using fluorescent tags, 
but the technique can only identify the position of well separated molecules; this modality provides up to several hundred snapshots covering a few seconds, yielding a few tens of trajectories.

Static images of receptors (even in the absence of ligand) typically reveal a {\em clustering} pattern, where receptors tend to accumulate in groups ranging from a few to a few tens of receptors. This occurs for VEGFR and also other receptor types, for which there is no evidence of a collective binding mechanism. The generally accepted explanation is that receptors accumulate in {\em microdomains}: small, physically delimited areas of the cell membrane that result from partitioning by actin filaments (elements of the cytoskeleton) or are formed by local aggregations of specific types of lipids and / or membrane proteins.
 However, there is no clear understanding of the mechanism of receptor accumulation. One hypothesis is that some microdomains have a specific molecular composition that results in an affinity for the receptors; receptors may diffuse in and out of them, but the crossing probability is asymmetric.
This hypothesis of {\em confining [micro]domains } is supported by time resolved tracking data, that reveals anomalous (non-Brownian) diffusion and under some circumstances, spatial confinement.

\subsubsection*{From microscopic details to global behavior}

%
Our model building program relies on a sequence of models, with three different levels of detail. Abstractions and/or average behaviors obtained from one level serve as inputs to the next, higher level. We use the idea of high affinity patches as a working hypothesis.

{\bf 1.~}At the {\em microscopic}  level, we investigate the localization, motion and interactions of {\em individual receptors}. Static distributions (Fig. \ref{fig:TEMillustration}), exhibit clusters that are not consistent with a random distribution. The identification of clusters can be done by a hierarchic clustering algorithm. The distribution of nearest neighbor distances, as well as other statistical measures, point toward a structure of {\em high density patches}, essentially identified with the observed clusters. Receptors are distributed randomly {\em within} the patches,
and the patches themselves also appear to be distributed randomly.

The observed receptor trajectories exhibit anomalous diffusion. We model this with {\em random walks} in the presence of various geometries of semi-permeable barriers. Comparisons of simulated and experimental step size distributions also support the high density patch hypothesis. In summary, the microscopic data combined with a Brownian motion model can provide estimations of the {\em individual and combined size}, as well as the {\em attractiveness} of the high density patches. In addition, direct measurements based on SPT can provide {\em exit and entrance rates} as well as {\em dimerization and dissociation rates} for molecular species of interest.

{\bf 2.~}The information on the size and properties of high affinity patches is used at the intermediate, {\em mesoscopic} level, to simulate the reactions and interchange of receptors and receptor-ligand complexes. At this level, each high density patch is abstracted into a single, well mixed compartment. Since receptors tend to diffuse quickly through the non-attractive region until they are [re]trapped by an attractive patch, the entire non-attractive region is represented as a single compartment.
\begin{figure}[!ht]
\captionsetup{font=scriptsize}
\centering
\includegraphics[trim= 0cm 12cm 0cm 0cm, clip=true, width=0.95\textwidth]{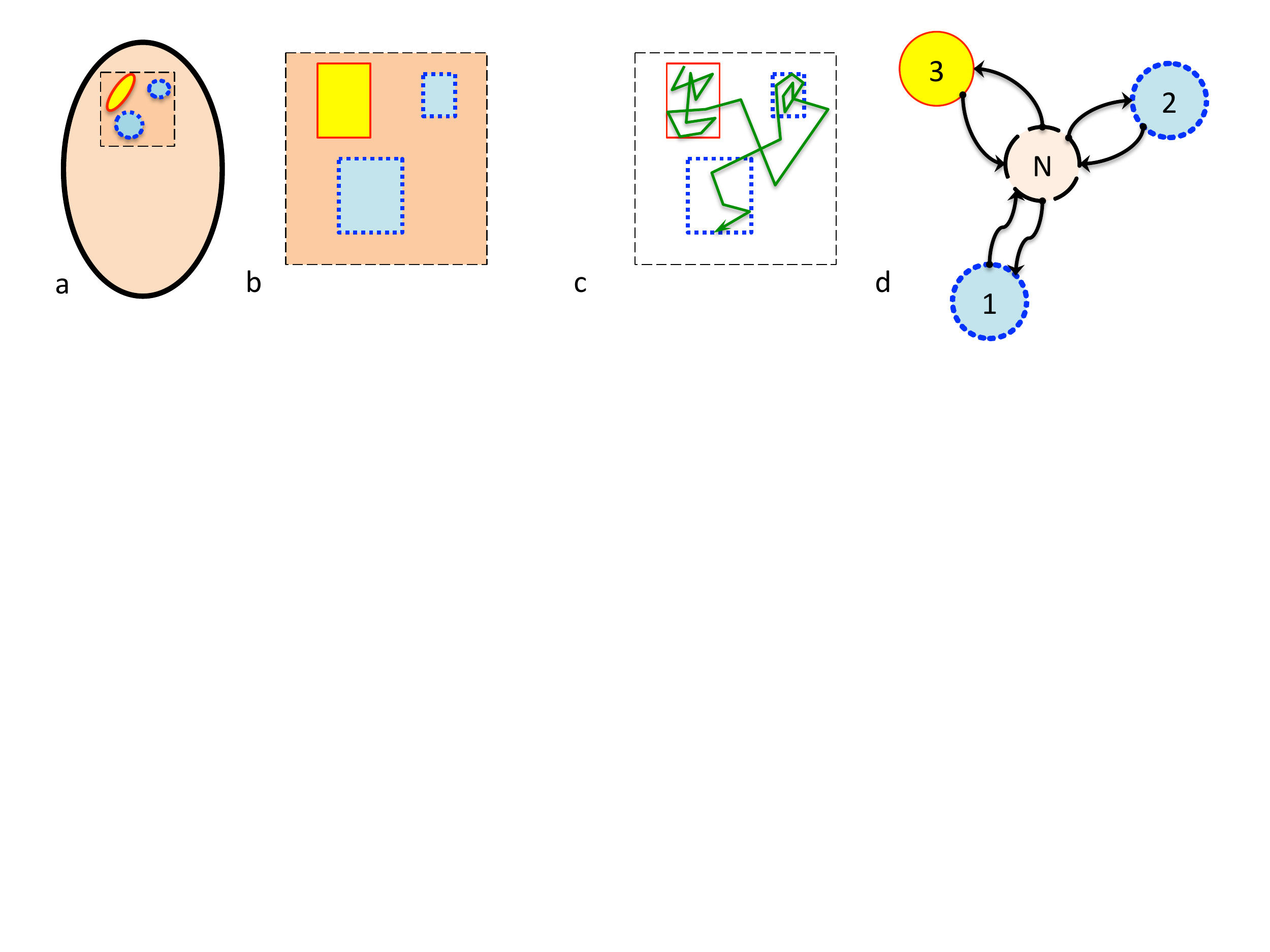}
\caption{
(a,b) Attractive microdomains occupy a small fraction of the cell membrane, and their measurements may be extracted from experimental images.
(c) Although the receptors can move through the non-attractive regions, they tend to remain in the smaller clusters. (d) In the mesoscopic approach each microdomain, as well as the rest of the membrane (the ``normal'' region), are represented as well-mixed compartments that may exchange particles.}
\end{figure}
The mesoscopic model is an abstraction of the microscopic models, where {\em spatial degrees of freedom have been discretized}. 
Mobility information is encapsulated in the particle exchange rates between domains, the capacity of the corresponding compartments, as well as the effective dimerization rates with the compartments. In terms of implementation, this level requires the composition of a spatial network of domains, defined by an oriented, weighted adjacency graph, and a chemical reaction network.

{\bf 3.~}The third, highest level of abstraction is obtained by (1) merging all attractive patches into a single one and (2) making the continuum approximation. The resulting {\em ODE system} is discussed in the remainder of this paper.

\section{Model}


Consistent with the emerging experimental picture, we make two assumptions. First, we assume that a fraction $f \le 1$ of the membrane is covered by domains that have a physical affinity for receptors. As receptors diffuse throughout the membrane, the probability of crossing the boundary of such an attractive domain is asymmetric - all else being equal, inbound crossing is $\alpha \ge 1$ times more likely than outbound. We will consider the aggregate of the high affinity patches as a single {\em high density} domain, 
 and refer to the rest 
 as the low density domain or sector. Second, we will set the mobility of receptor dimers lower than that of monomers.
These two ingredients result in the preferential accumulation of receptors in the high density patches. We investigate the effect of this accumulation (clustering) on dimerization and signal initiation. We are especially interested in establishing whether there is a postive feed-back between dimerization and receptor clustering.


\subsection{Reactions and Equations}

\begin{figure*}
\captionsetup{font=scriptsize}
\begin{center}
\includegraphics[scale=0.4]{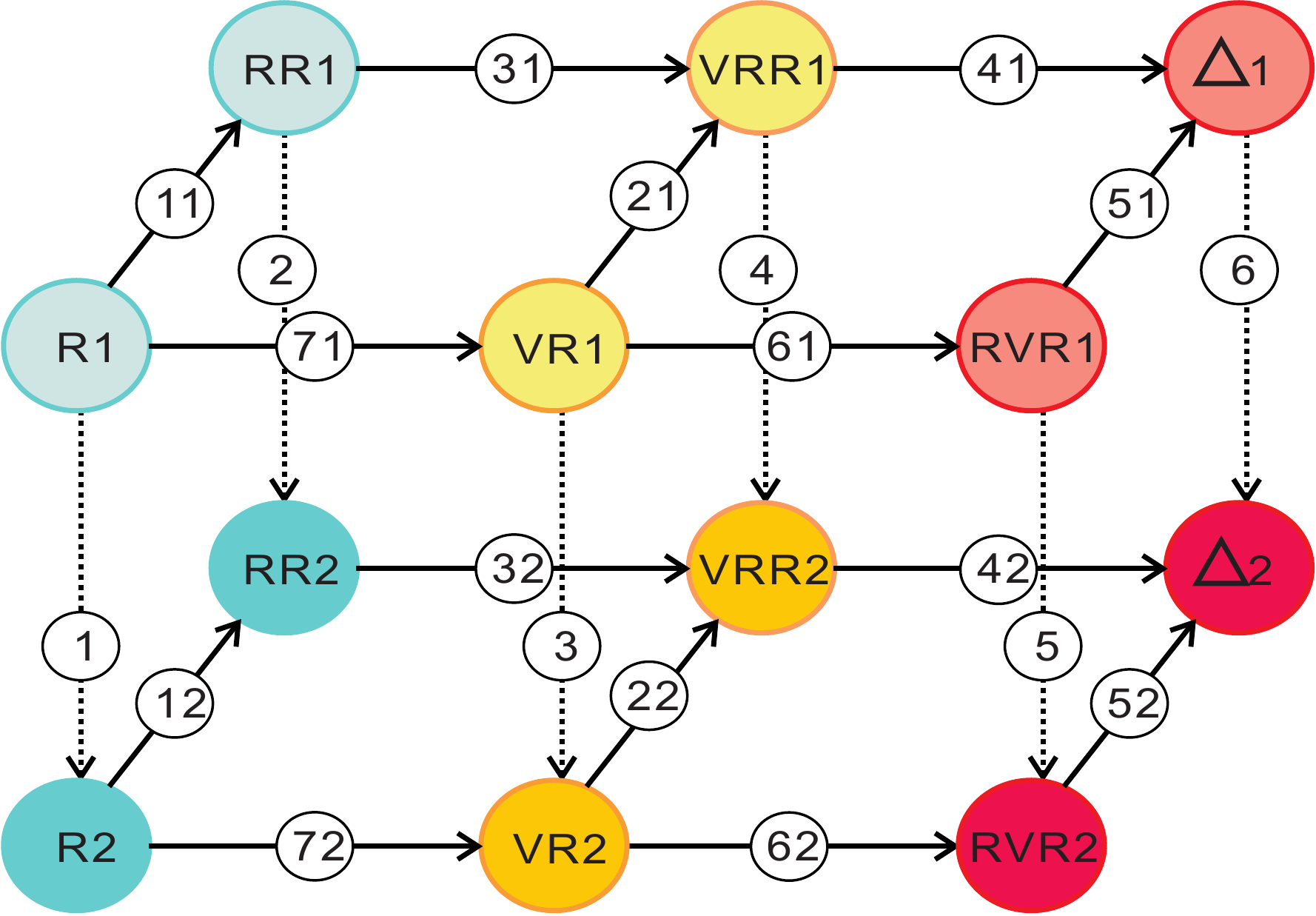}
\vspace{-0.5cm}
\end{center}
\caption{Reactions in the two-compartment model. Each horizontal sheet contains the reactions in one of the domains; transport reactions are "vertical". Here we omitted the added monomer receptors in reactions $\mathcal{C}_\mathrm{1x}$ and $\mathcal{C}_\mathrm{2x}$, as well as the added VEGF ($V$) in reactions $\mathcal{C}_\mathrm{3x}$ and $\mathcal{C}_\mathrm{7x}$. }
\label{allreact}
\end{figure*}

We follow the mathematical modeling framework of MacGabhann, Popel and coworkers \cite{MacGabhannP2007} to describe free ($\rmono$) and ligand-bound ($\vr$) receptors, receptor dimers ($\rr$), and three ligand-bound dimer complexes ($\vrr$, $\rvr$, $\Delta$); the ligand is considered constant. Their structure and reactions among them are illustrated in Figure \ref{7react}. For simplification, we assume that there is a region in the membrane with high affinity for VEGF receptors, and describe the rest as a second,``normal'' or low affinity one. Each of the six species is presented in both domains; similary, each of the 7 reactions has a copy in each domain, see Figure \ref{allreact}.
%
Assuming the free VEGF concentration is kept constant at $V_0$, we have a 12-dimensional state vector,
\begin{equation}
\mathbf{X} = \left( \mathrm{[R_1], [R_2], [RR_1], [RR_2], [VR_1], [VR_2],[VRR_1], [VRR_2], [RVR_1], [RVR_2], [\Delta_1], [\Delta_2]}\right)^T~~.
\end{equation}
In addition to the 28 (irreversible) reactions that represent molecular transformations, we describe the transfer of every molecular species between domains as a separate reaction, bringing the total to 40 (irreversible) reactions.  It is convenient to group pairs of opposing reactions into single reversible reactions \cite{KleimanMCLS2011}, leaving us with 20 reversible reactions, 
 as illustrated in Figure \ref{allreact}.
The arrows represent the conventional direction for the corresponding fluxes. The 20 reactions with reaction rates assuming mass-action are denoted as follows (where $x=1,2$):
\begin{eqnarray}
\begin{array}{llll}
\mathrm{\mathcal{C}_{1x}:~} \rmono\x + \rmono\x \xrightleftharpoons[d]{b}  \rr\x   &
\mathrm{\mathcal{C}_{2x}:~}\vr\x + \rmono\x \xrightleftharpoons[d]{b}  \vrr\x; &
\mathrm{\mathcal{C}_{3x}:~} \rr\x + \vc _0 \xrightleftharpoons[c]{2a}  \vrr\x  ;\\
\mathrm{\mathcal{C}_{4x}:~} \vrr\x \xrightleftharpoons[c_{i}]{a_{i}} \Delta\x;   &
\mathrm{\mathcal{C}_{5x}:~} \rvr\x \xrightleftharpoons[d_{i}]{b_{i}} \Delta\x ; &
\mathrm{\mathcal{C}_{6x}:~} \vr\x + \rmono\x \xrightleftharpoons[c]{a_{s}} \rvr\x;\\
& \mathrm{\mathcal{C}_{7x}:~} \rmono\x + \vc _0 \xrightleftharpoons[c]{a} \vr\x  ;  &\\
\mathrm{\mathcal{D}_{1}:~} \rmono_1\xrightleftharpoons[k_2]{k_1} \rmono_2; &
\mathrm{\mathcal{D}_{2}:~} \rr_1\xrightleftharpoons[\beta k_2]{\beta k_1} \rr_2   ; &
\mathrm{\mathcal{D}_{3}:~} \vr_1\xrightleftharpoons[k_2]{k_1} \vr_2;& \\
\mathrm{\mathcal{D}_{4}:~} \vrr _1\xrightleftharpoons[\beta k_2]{\beta k_1} \vrr_2   ; &
\mathrm{\mathcal{D}_{5}:~} \rvr_1\xrightleftharpoons[\beta k_2]{\beta k_1} \rvr_2; &
\mathrm{\mathcal{D}_{6}:~} \Delta_1\xrightleftharpoons[\beta k_2]{\beta k_1} \Delta_2. &
\end{array}
\label{eq:AllReactions}
\end{eqnarray}
The corresponding stoichiometry matrix is
\begin{displaymath}
\scalemath{0.7}{\Gamma =
\left[ \begin{array}{ccccccccccccccccccccccccccc}
-2	&	0	&	-1	&	0	&	0	&	0	&	0	&	0	&	0	&	0	&	-1	&	0	&	-1	&	0	&	-1	&	0	&	0	&	0	&	 0	&	 0	\\
0	&	-2	&	0	&	-1	&	0	&	0	&	0	&	0	&	0	&	0	&	0	&	-1	&	0	&	-1	&	1	&	0	&	0	&	0	&	 0	&	 0	\\
1	&	0	&	0	&	0	&	-1	&	0	&	0	&	0	&	0	&	0	&	0	&	0	&	0	&	0	&	0	&	-1	&	0	&	0	&	 0	&	 0	\\
0	&	1	&	0	&	0	&	0	&	-1	&	0	&	0	&	0	&	0	&	0	&	0	&	0	&	0	&	0	&	1	&	0	&	0	&	 0	&	 0	\\
0	&	0	&	-1	&	0	&	0	&	0	&	0	&	0	&	0	&	0	&	-1	&	0	&	1	&	0	&	0	&	0	&	-1	&	0	&	 0	&	 0	\\
0	&	0	&	0	&	-1	&	0	&	0	&	0	&	0	&	0	&	0	&	0	&	-1	&	0	&	1	&	0	&	0	&	1	&	0	&	 0	&	 0	\\
0	&	0	&	1	&	0	&	1	&	0	&	-1	&	0	&	0	&	0	&	0	&	0	&	0	&	0	&	0	&	0	&	0	&	-1	&	 0	&	 0	\\
0	&	0	&	0	&	1	&	0	&	1	&	0	&	-1	&	0	&	0	&	0	&	0	&	0	&	0	&	0	&	0	&	0	&	1	&	 0	&	 0	\\
0	&	0	&	0	&	0	&	0	&	0	&	0	&	0	&	-1	&	0	&	1	&	0	&	0	&	0	&	0	&	0	&	0	&	0	&	 -1	&	 0	\\
0	&	0	&	0	&	0	&	0	&	0	&	0	&	0	&	0	&	-1	&	0	&	1	&	0	&	0	&	0	&	0	&	0	&	0	&	 1	&	 0	\\
0	&	0	&	0	&	0	&	0	&	0	&	1	&	0	&	1	&	0	&	0	&	0	&	0	&	0	&	0	&	0	&	0	&	0	&	 0	&	 -1	\\
0	&	0	&	0	&	0	&	0	&	0	&	0	&	1	&	0	&	1	&	0	&	0	&	0	&	0	&	0	&	0	&	0	&	0	&	 0	&	 1	\\
\end{array} \right].}
\end{displaymath}
\paragraph{Effective concentrations:}
We will use {\em effective concentrations} to describe the amounts of each species found in the two domains; $\mathrm{[S_x]^{eff}}$ is defined as the ratio of the amount (number of mols) of substance $S$ in domain $x$ ($x=1,2$), divided by the {\em total} area of the cell membrane $A_\mathrm{cell}$. We will refer to the usual concentrations as {\em physical}, $\mathrm{[S_x]^{phys}}$. Generally, the meaning of the concentrations and rate constants is similar to the standard approach in \cite{MacGabhannP2007}, with some important differences as discussed below.

\begin{figure}
\captionsetup{font=scriptsize}
\begin{center}
\includegraphics[scale=0.5]{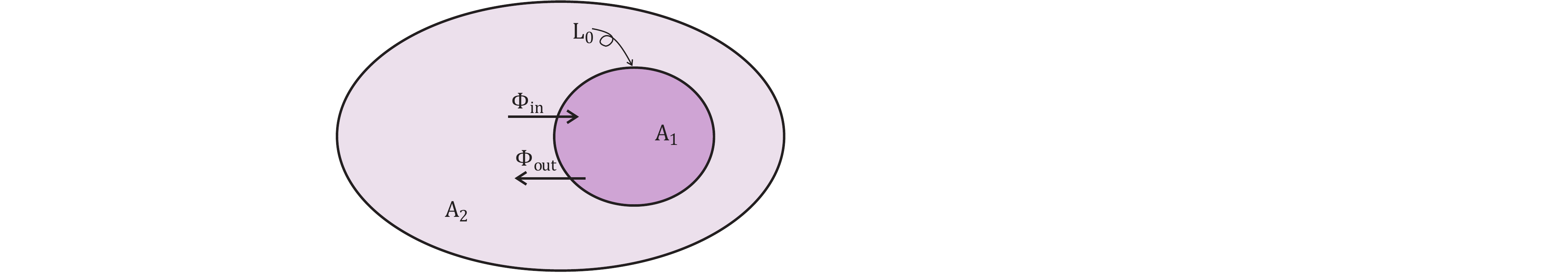}
\end{center}
\caption{Schematic and notations for the high- and low-density areas on the cell surface. We assume that a fraction of the membrane (area $A_1$) has a higher affinity for VEGF receptors than the rest of the membrane. This translates into asymmetric rate 'constants' for the $\Phi_\mathrm{in}$, and $\Phi_\mathrm{out}$ fluxes.}
\label{cell}
\end{figure}

Consider first the exchange reactions ($\mathcal{D}_1 \ldots \mathcal{D}_6$ in eq.(\ref{eq:AllReactions}) ), exemplified by reaction $\mathcal{D}_\mathrm{1}: \rmono_1\xrightleftharpoons[]{} \rmono_2$. 
Let the fraction of the area that has high affinity to VEGF receptors be $f$. The size of the high (VEGF) density area is $A_1=f\cdot A_\mathrm{cell}$, and the remaining area is $A_2=(1-f)\cdot A_\mathrm{cell}$ (see Figure \ref{cell}).
%
Let us derive the flux of unbound receptors $\rmono$ between $A_1$ and $A_2$, represented by the reaction $\rmono_1\xrightleftharpoons[]{} \rmono_2$. 
Let $\mathrm{[R_1]^{phys}}$ and $\mathrm{[R_2]^{phys}}$ be the physical concentrations of $\rmono$ in $A_1$ and $A_2$, respectively, 
defined as the amount (in fmol) of $\rmono$ in $A_1$ (respectively $A_2$), divided by the area $A_1$ (resp. $A_2$). The effective concentrations, denoted as $\mathrm{[R_1]^{eff}}$ ($\mathrm{[R_2]^{eff}}$ resp.), are the amounts of $\rmono$ in $A_1$ ($A_2$ resp.), but divided by the {\em total} area $A_\mathrm{cell}$. Therefore,
\begin{equation}
  \mathrm{[R_1]^{phys}}=\frac{\mathrm{[R_1]^{eff}}}{f}~~~\mathrm{ and}   ~~~\mathrm{[R_2]^{phys}}=\frac{\mathrm{[R_2]^{eff}}}{(1-f)}~~~,
\label{eq:EffConDef}
\end{equation}
 with units of $\mathrm{ fmol / cm^{2}}$ for all concentrations.

We assume that the {\em flux} of receptors,  $\Phi_\mathrm{out}$ (amount of substance per unit time and boundary length, in $\mathrm{fmol / (cm \cdot s)}$ in our case), from $A_1$ to $A_2$ is proportional to the {\em physical} concentration, $\Phi_\mathrm{out} = \gamma_\mathrm{out} \mathrm{[R_1]^{phys}}$; similarly, the receptor flux into $A_1$ is $\Phi_\mathrm{in} = \gamma_\mathrm{in} \mathrm{[R_2]^{phys}}$. The factors $\gamma_\mathrm{in}$ and $\gamma_\mathrm{out}$ reflect the physical permeability of the boundary and have units of $\mathrm{cm / s}$. We define the {\em attractiveness}, $\alpha \equiv \gamma_\mathrm{in} / \gamma_\mathrm{out}$, to reflect the asymmetry of the permeabilities; so we have $\Phi_\mathrm{in} = \gamma_\mathrm{in} [R_2]^{phys}=\alpha\gamma_\mathrm{out} [R_2]^{phys}$, and $\alpha \ge 1$ means that a receptor or a dimer is more easily transferred into the high concentration area $A_1$ than into $A_2$. Consequently, the exchange fluxes between the two domains will balance when the ratio of the respective {\em physical} concentrations is $\alpha$, i.e. $\Phi_\mathrm{in}=\Phi_\mathrm{out} ~\Leftrightarrow~ \alpha \gamma_\mathrm{out} \mathrm{[R_2]^{phys}} = \gamma_\mathrm{out} \mathrm{[R_1]^{phys}} ~\Leftrightarrow~ \alpha \mathrm{[R_2]^{phys}} =  \mathrm{[R_1]^{phys}} $.

Consider the net rate of change of concentrations $\mathrm{[R_1]^{phys}}$ and $\mathrm{[R_2]^{phys}}$, due to the exchange of receptor monomers between the two compartments, we have
\begin{eqnarray} \label{EqnCon1}
\left( \frac{d\mathrm{[R_1]^{phys}}}{dt} \right)_{\Phi}= \frac{(\Phi_\mathrm{in}-\Phi_\mathrm{out}) \cdot L_0}{A_1}~~;~~
\left( \frac{d\mathrm{[R_2]^{phys}}}{dt} \right)_{\Phi}= \frac{(\Phi_\mathrm{out}-\Phi_\mathrm{in}) \cdot L_0}{A_2},
\end{eqnarray}
where $L_0$ is the length of the boundary between $A_1$ and $A_2$. Substitute  $A_1$, $A_2$, $\Phi_1$ and $\Phi_2$ into (\ref{EqnCon1}):
\begin{eqnarray} \label{EqnCon3}
\left( \frac{d\mathrm{[R_1]^{phys}}}{dt} \right)_{\Phi} =  (\alpha \mathrm{[R_2]^{phys}}-\mathrm{[R_1]^{phys}})
\frac{L_0 \gamma_\mathrm{out} }{f A_\mathrm{cell}}~~,~~
\left( \frac{d\mathrm{[R_2]^{phys}}}{dt} \right)_{\Phi} =
(\mathrm{[R_1]^{phys}}-\alpha\mathrm{[R_2]^{phys}}) \frac{L_0 \gamma_\mathrm{out} } {(1-f)A_\mathrm{cell}}.
\end{eqnarray}
Defining a common time constant $\delta \equiv A_\mathrm{cell} / (L_0\gamma_\mathrm{out})$ in (\ref{EqnCon3}),
we have
\begin{eqnarray} \label{EqnCon5}
\left( \frac{d\mathrm{[R_1]^{phys}}}{dt} \right)_{\Phi} =(\alpha\mathrm{[R_2]^{phys}}-\mathrm{[R_1]^{phys}})\frac{1}{f\delta}~~,~~
\left( \frac{d\mathrm{[R_2]^{phys}}}{dt} \right)_{\Phi} =(\mathrm{[R_1]^{phys}}-\alpha\mathrm{[R_2]^{phys}})\frac{1}{(1-f)\delta}.
\end{eqnarray}
Finally, substituting the effective concentrations from (\ref{eq:EffConDef})
yields
\begin{eqnarray*}
\left( \frac{d\mathrm{[R_1]^{eff}}}{dt}  \right)_{\Phi} &=&\frac{\alpha}{\delta(1-f)}\mathrm{[R_2]^{eff}}-\frac{1}{\delta f}\mathrm{[R_1]^{eff}},
\\
\left( \frac{d\mathrm{[R_2]^{eff}}}{dt}  \right)_{\Phi} &=&-\frac{\alpha}{\delta(1-f)}\mathrm{[R_2]^{eff}}+\frac{1}{\delta f}\mathrm{[R_1]^{eff}}.
\label{eq:ExchFluxesEff}
\end{eqnarray*}
The above result implies the identity  $\frac{d\mathrm{[R_1]^{eff}}}{dt}+\frac{d\mathrm{[R_2]^{eff}}}{dt}=0$, which reflects particle number conservation. This is the main advantage of using effective concentrations.
%

We follow the same line of reasoning for the other transfer reactions.
The exit rate constants also reflect the generic mobility of particles; in a more detailed simulation, one could relate them to the diffusion constants and the permeability of the membranes. Here we will assume that exit rate constants are the same as above for ligand-bound monomer species $VR_x$. 
 For dimer species, $RR_x$, $VRR_x$, $RVR_x$ and $\Delta_x$, we will use reduced exit rate constants, proportional to $\gamma_\mathrm{in}$ and $\gamma_\mathrm{out}$, and denote the coefficient as $\beta$. In summary, the six exchange fluxes are
\begin{displaymath}
\begin{array}{llll}
& \phi_1=k_1\mathrm{[R_1]^{eff}}-k_2\mathrm{[R_2]^{eff}},\quad
& \phi_2=\beta(k_1\mathrm{[RR_1]^{eff}}-k_2\mathrm{[RR_2]^{eff}}),\\
& \phi_3=k_1\mathrm{[VR_1]^{eff}}-k_2\mathrm{[VR_2]^{eff}},\quad
& \phi_4=\beta(k_1\mathrm{[VRR_1]^{eff}}-k_2\mathrm{[VRR_2]^{eff}}),\\
& \phi_5=\beta(k_1\mathrm{[RVR_1]^{eff}}-k_2\mathrm{[RVR_2]^{eff}}),\quad
& \phi_6=\beta(k_1\mathrm{[\Delta_1]^{eff}}-k_2\mathrm{[\Delta_2]^{eff}}),\\
\end{array}
\end{displaymath}
where $k_1=\frac{1}{\delta f}$ and $k_2=\frac{\alpha}{\delta(1-f)}$.

Next, we consider the rates of chemical reactions, molecular transformations that take place within each area. As an example, consider a reaction in the high density area.  For  $\mathcal{C}_\mathrm{21}:\vr_1 + \rmono_1 \xrightleftharpoons[d]{b}  \vrr_1$, we have \begin{eqnarray} \label{EqnCon7}
\frac{d\mathrm{[ R_1]^{phys}}}{dt}=\frac{\mathrm{[ dVR_1]^{phys}}}{dt}&=& -b\mathrm{[ R_1]^{phys}}\mathrm{[ VR_1]^{phys}}+d\mathrm{[ VRR_1]^{phys}},\nonumber
\\
\label{EqnCon8}
\label{eq:RateLawPhysConc}
\frac{d\mathrm{[ VRR_1]^{phys}}}{dt}&=& b\mathrm{[ R_1]^{phys}}\mathrm{[ VR_1]^{phys}}-d\mathrm{[ VRR_1]^{phys}}.
\end{eqnarray}
Substituting $\mathrm{[ R_1]^{phys}}=\mathrm{[ R_1]^{eff}} /f $, $\mathrm{[ VR_1]^{phys}}=\mathrm{[ VR_1]^{eff}}/ f$ and $\mathrm{[ RVR_1]^{phys}}=\mathrm{[ RVR_1]^{eff}} / f$, we have
 \begin{eqnarray}
\frac{d\mathrm{[ VRR_1]^{eff}}}{dt}=-\frac{d\mathrm{[ R_1]^{eff}}}{dt}=-\frac{d\mathrm{[ VR_1]^{eff}}}{dt}
= \frac{b}{f}\mathrm{[ R_1]^{eff}}\mathrm{[ VR_1]^{eff}}-d\mathrm{[ VRR_1]^{eff}}~~
,
\label{eq:RateLawEffConc}
\end{eqnarray}
therefore the flux for reaction $r_{21}$ is
\[\phi_{21}=\frac{b}{f}\mathrm{[ R_1]^{eff}}\mathrm{[ VR_1]^{eff}}-d\mathrm{[ VRR_1]^{eff}}~~.\]
The only difference between the above rate law and the one in terms of physical concentrations (\ref{eq:RateLawPhysConc}), is that the dimerization rate constant is scaled by the relative size of the domain, $b \rightarrow b / f$. This reflects the effect of clustering on dimerization; if the same number of reacting molecules are forced into a smaller space, their collision rate and implicitly, the absolute dimerization rate, will increase. Similar considerations give the following for the 14 reversible reactions:
\begin{eqnarray}
\begin{array}{ll}
\phi_{11} = \frac{2b}{f} \mathrm{[ R_1]^{eff}}^\mathrm{2}-d_1\mathrm{[ RR_1]^{eff}}, & \phi_{12}=\frac{2b}{1-f}\mathrm{[ R_2]^{eff}}^\mathrm{2}-d\mathrm{[ RR_2]^{eff}} \\
\phi_{21} = \frac{b}{f} \mathrm{[ R_1]^{eff}}\mathrm{[ VR_1]^{eff}}-d\mathrm{[ VRR_1]^{eff}}, &\phi_{22}=\frac{b}{1-f}\mathrm{[ R_2]^{eff}}\mathrm{[ VR_2]^{eff}}-d_2\mathrm{[ VRR_2]^{eff}} \\
\phi_{31} = 2aV_0\mathrm{[ RR_1]^{eff}}-c\mathrm{[ VRR_1]^{eff}}, & \phi_{32}=2aV_0\mathrm{[ RR_2]^{eff}}-c\mathrm{[ VRR_2]^{eff}} \\
\phi_{41} = a_{i}\mathrm{[ VRR_1]^{eff}}-2c_{i}\mathrm{[ \Delta_1]^{eff}}, & \phi_{42}=a_{i}\mathrm{[ VRR_2]^{eff}}-2c_{i}\mathrm{[ \Delta_2]^{eff}} \\
\phi_{51} = b_{i}\mathrm{[ RVR_1]^{eff}}-d_{i}\mathrm{[ \Delta_1]^{eff}}, & \phi_{52}=b_{i}\mathrm{[ RVR_2]^{eff}}-d_{i}\mathrm{[ \Delta_2]^{eff}} \\
\phi_{61} = \frac{a_{s}}{f} \mathrm{[ R_1]^{eff}}\mathrm{[ VR_1]^{eff}}-c\mathrm{[ RVR_1]^{eff}}, & \phi_{62}=\frac{a_{s}}{1-f}\mathrm{[ R_2]^{eff}}\mathrm{[ VR_2]^{eff}}-c\mathrm{[ RVR_2]^{eff} }\\
\phi_{71} = aV_0\mathrm{[ R_1]^{eff}}-c\mathrm{[ VR_1]^{eff}}, & \phi_{72}=aV_0\mathrm{[ R_2]^{eff}}-c\mathrm{[ VR_2]^{eff}}~~.
\end{array}
\end{eqnarray}

We denote by $\Phi(X)=\left( \phi_{11}, \phi_{12}, \phi_{21}, \cdots, \phi_{71},\phi_{72},\phi_1,\cdots,\phi_6 \right)^T$, then the system of differential equation assuming mass-action is
\begin{equation}
\label{ODE}
\frac{d \mathbf{X} }{dt}=\Gamma \cdot \Phi( \mathbf{X} ).
\end{equation}

\subsection{Parameter values}

\begin{figure}[h!]
\begin{center}
\vspace{-0.2in}
\includegraphics[trim=0.4in 0.3in 0.4in 0.25in, clip, width=0.4\textwidth]{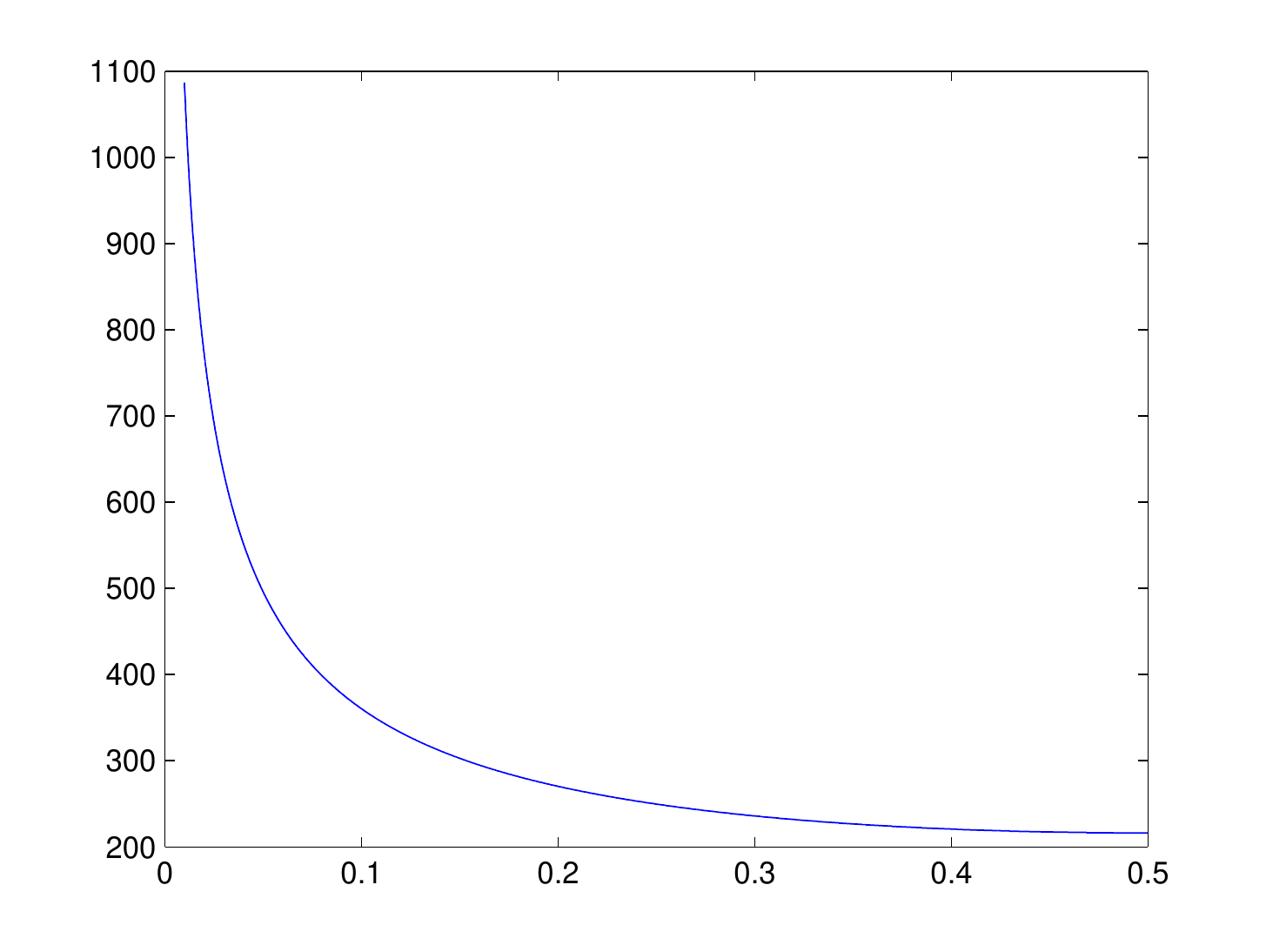}
\end{center}
\vspace{-0.3in}
\caption{Graph of $\delta$ as a function in $f$ ($0\le f\le 0.5$).}
\label{deltaf}
\vspace{-0.2in}
\end{figure}

\paragraph{Receptor Diffusivity and Boundaries: }
For the diffusivity of VEGFR, we used the exit rate of $\gamma_\mathrm{out}=8.23\cdot 10^{-6} ~\mathrm{cm / s}$
 based on the expression given in \cite{MacGabhannP2007}. 
The cells are assumed to have a surface area of $1000 \mathrm{ \mu m^2}$ \cite{MacGabhannP2007}; assuming a spherical shape,
 the radius of a cell works out to approximately  $r_{cell}=8.9 ~\mathrm{\mu m}$, and the length of the high density area boundary is $L_0=2\pi \sqrt{r_{cell}^2-\left({r_{cell}-\frac{1000f}{2\pi\cdot r_{cell}}}\right)^2}  \mathrm{ \mu m}$, where $f \equiv A_1 / A_\mathrm{cell}$ is the relative fraction of the HD area. The $L_0$ and $\gamma_\mathrm{out}$ can be readily substituted into the definition 
 $\delta = A_\mathrm{cell} / (L_0 \gamma_\mathrm{out})$.
The graph of $\delta$ as a function of $f$ is shown in Figure \ref{deltaf}. We can see that $\delta$ decreases faster in the beginning as $f$ is increasing from 0 to 0.5.

\noindent
Following \cite{MacGabhannP2007}, we assume that there are 40,000 receptor monomers on one cell membrane, corresponding to $R_\mathrm{total} \approx 6.6~ \mathrm{fmol\cdot cm^{-2}}$. 

\paragraph{Reaction Rates:}
We use the model of \cite{MacGabhannP2007}, with base units as follows: volume concentration (of VEGF only, $V_0$), in $\mathrm{nM}$; all surface concentrations in $\mathrm{fmol / cm^2}$; time in $\mathrm{s}$. Except for VEGF, whose concentration is {\em not} a variable, all molecular species in the model are surface bound. Consequently, the units of mass-action rate constants are $\mathrm{s^{-1}}$ for unimolecular reactions, $ \mathrm{( nM \cdot s)^{-1}}$ for reactions involving VEGF, and $ \mathrm{ cm^2 / (fmol \cdot s) }$ for on-surface dimerizations. The rate constants are summarized in the table below.
\vspace{0.5cm}

\scalemath{0.7}{
\begin{tabular}{c c c c c |c c c c c c}
\hline
{\bf Reaction} & {\bf Direction} & {\bf Notation} & {\bf Value} & {\bf Unit } & {\bf Reaction }& {\bf Direction }& {\bf Notation }& {\bf Value }& {\bf Unit }\\
\hline
$\rmono\x + \rmono\x \xrightleftharpoons[]{}  \rr\x   $ & $ \rightarrow$ & $b$ & 0.1 & $\mathrm{cm^2 / (fmol\cdot s)}$ & $\rmono_1\xrightleftharpoons[]{} \rmono_2$ & $ \rightarrow$ &$k_1$ & 0.0277&  $\mathrm{s^{-1}}$\\
                               & $ \leftarrow$ & $d$  & 0.01 & $\mathrm{s^{-1}}$ &                      & $ \leftarrow$  &$k_2$ & 0.0154& $\mathrm{s^{-1}}$ \\
\hline
$\vr\x + \rmono\x \xrightleftharpoons[]{}  \vrr\x$ & $ \rightarrow$ & $b$ & 0.1 & $\mathrm{cm^2 / (fmol\cdot s)}$ & $\rr_1\xrightleftharpoons[]{} \rr_2$ & $ \rightarrow$ &$\beta k_1$ & 0.01385 &  $\mathrm{s^{-1}}$\\
                               & $ \leftarrow$ & $d$  & 0.01 & $\mathrm{s^{-1}}$  &                     & $ \leftarrow$  &$\beta k_2$ & 0.0077&  $\mathrm{s^{-1}}$\\
\hline
$\rr\x + \vc_0 \xrightleftharpoons[]{}  \vrr\x$ & $ \rightarrow$ & $2a$ & 0.0044 & $\mathrm{ (nM \cdot s)^{-1}}$ & $\vr_1\xrightleftharpoons[]{} \vr_2$ & $ \rightarrow$ &$k_1$ &  0.0277&  $\mathrm{s^{-1}}$\\
                               & $ \leftarrow$ & $c$   & 0.026 & $\mathrm{s^{-1}}$ &                   & $ \leftarrow$  &$k_2$ &0.0154 &  $\mathrm{s^{-1}}$\\
\hline
$\vrr\x \xrightleftharpoons[]{} \Delta\x$ & $ \rightarrow$ & $a_{i}$ & 0.949 & $\mathrm{s^{-1}}$ &$\vrr_1\xrightleftharpoons[]{} \vrr_2$ & $ \rightarrow$ &$\beta k_1$ &  0.01385&  $\mathrm{s^{-1}}$\\
                               & $ \leftarrow$ & $c_{i}$& 0.026 & $\mathrm{s^{-1}}$ &                      & $ \leftarrow$  &$\beta k_2$ & 0.0077&  $\mathrm{s^{-1}}$\\
\hline
$\rvr\x \xrightleftharpoons[]{} \Delta\x$ & $ \rightarrow$ & $b_{i}$ & 0.446 & $\mathrm{s^{-1}}$ &$\rvr_1\xrightleftharpoons[]{} \rvr_2$ & $ \rightarrow$ &$\beta k_1$ &  0.01385&  $\mathrm{s^{-1}}$\\
                               & $ \leftarrow$ & $d_{i}$& 0.02 &  $\mathrm{s^{-1}}$ &                    & $ \leftarrow$  &$\beta k_2$ &0.0077 &  $\mathrm{s^{-1}}$\\
\hline
$\vr\x + \rmono\x \xrightleftharpoons[]{} \rvr\x$ & $ \rightarrow$ & $a_{s}$ & 0.21 & $\mathrm{cm^2 / (fmol\cdot s)}$ &$\Delta_1\xrightleftharpoons[]{} \Delta_2$ & $ \rightarrow$ &$\beta k_1 $&  0.01385&  $\mathrm{s^{-1}}$\\
                               & $ \leftarrow$ & $c$& 0.026 &  $\mathrm{s^{-1}}$  &              & $ \leftarrow$  &$\beta k_2 $&0.0077 &  $\mathrm{s^{-1}}$\\
\hline
$\rmono\x + \vc_0 \xrightleftharpoons[]{} \vr\x$ & $ \rightarrow$ & $a$ & 0.0044 &$\mathrm{ (nM \cdot s)^{-1}}$&  \multicolumn{5}{l}{NOTE: the values for $k_1$ and $k_2$ given here correspond to}  \\
                               & $ \leftarrow$ & $c$& 0.026 &$\mathrm{s^{-1}}$&  \multicolumn{5}{l}{ $f=0.1$, $\alpha=5$, $\beta=0.5$.} \\
\hline
\end{tabular}}

\subsection{Steady States}

To solve for the closed form steady-state of the differential equation system (\ref{ODE}), we use the method introduced in \cite{HalaszLMRE2013}.
The steady states of this system are sets of concentration values $\{ \mathrm{ [R_1], [R_2], [RR_1], [RR_2]}$ $\mathrm{ [VR_1], [VR_2], [VRR_1], [VRR_2], [RVR_1], [RVR_2], [\Delta_1], [\Delta_2]} \}$ for which the expression on the right-hand side of (\ref{ODE}) is identically zero. 
Define an expanded vector $\bX_E$ that consists of the original variables of $\bX$ plus the binomials $\mathrm{ [R_1]^2} $, $\mathrm{[R_2]^2}$, $\mathrm{[R_1]\cdot[VR_1]} $ and $\mathrm{[R_2]\cdot[VR_2]}$ as
\begin{eqnarray}
\bX_E \equiv\left(\mathrm{ [R_1], [R_2], [RR_1], [RR_2] [VR_1], [VR_2],
[VRR_1], [VRR_2], [RVR_1], [RVR_2],
} \right. \nonumber\\ \left. \mathrm{
 [\Delta_1], [\Delta_2], [R_1]^2, [R_2]^2, [R_1]\cdot[VR_1], [R_2]\cdot[VR_2] } \right)^T~~.
\end{eqnarray}
As the rate law vector $\Phi(\bX)=\left( \phi_{11}, \phi_{12}, \phi_{21}, \cdots, \phi_{71},\phi_{72},\phi_1,\cdots,\phi_6 \right)^T$ is a linear combination of $\bX_E$, it can be interpreted as a linear expression:
$\Phi(\bX) = \bA_E \cdot \bX_E$. So we have
\[\frac{d \bX}{dt}=\Gamma\cdot\Phi(\bX)=\Gamma \cdot \bA_E \cdot \bX_E=\bar{\bA}_E \cdot \bX_E,\]
where $\bar{\bA}_E=\Gamma \cdot \bA_E$ (the  $12 \times 16$ dimensional expanded system matrix $\bar{\bA}_E$ is too large to reproduce within normal text). We substitute $X_1\equiv\mathrm{[R_1]^2}$, $X_2\equiv\mathrm{[R_2]^2}$, $Y_1\equiv\mathrm{[R_1] \cdot [VR_1]}$ and $Y_2\equiv\mathrm{[R_2]\cdot[VR_2]}$ into $\bX_E$, and denote the new vector as $\bar{\bX}_E$. Then all elements in $\bar{\bX}_E$ are linear variables, and the steady state problem is equivalent to that for a linear system $\frac{d\bX}{dt}=\Gamma \cdot \Phi(\bX) = \bar{\bA}_E \cdot \bar{\bX}_E$, find the set of $\bar{\bX}_E$ such that
\begin{equation}
\frac{d \bX}{dt}=0 ~\Leftrightarrow~ \bar{\bA}_E \cdot \bar{\bX}_E=0.
\label{ODEe}
\end{equation}

By Theorem 1 in \cite{HalaszLMRE2013}, $\bar{\bA}_E$ has the same rank as the original system, i.e. $rank(\bar{\bA}_E)=rank(\Gamma) = 11$. For the linear equation system $\bar{\bA}_E \cdot \bar{\bX}_E=0$, as $rank(\bar{\bA}_E) = 11$,  and there are 16 variables in $\bar{\bX}_E$, we can solve 11 variables (dependent variables) as a function of the other 5 (free variables). To achieve that, we first discard a row of $\bar{\bA}_E$, whose loss would not reduce the rank of $\bar{\bA}_E$. In this case, we select row 1.

Next, we select the 11 dependent variables. The set of dependent variables has to be determined carefully to make the method given in \cite{HalaszLMRE2013} work. We select
$\bar{\bX}_D=( \mathrm{ [RR_1],[RR_2],[VR_1],[VR_2],[VRR_1],[VRR_2],} $
$\mathrm{ [RVR_1],[RVR_2],[\Delta_1],[\Delta_2], }Y_2 )^T$
as the set of dependent variables, and $\bar{\bX}_F=( \mathrm{ [R_1],[R_2],} X_1, X_2, Y_1 )^T$ as the set of free variables. We use Cramer's Rule to solve for $\bar{\bX}_D$ in terms of $\bar{\bX}_F$. Denote the solution as
\[y_i=a_{i1}\mathrm{[R_1]}+a_{i2}\mathrm{[R_2]}+a_{i3}X_1+a_{i4}X_2+a_{i5}Y_1,\]
 where $i=1,2,\cdots, 11$, $y_i\in X_D$ and $a_{ij}$ are algebraic combinations of reaction rate constants. Substitute the bilinears $X_1=\mathrm{[R_1]}^2$, $X_2=\mathrm{[R_2]}^2$, $Y_1=\mathrm{[R_1]}\mathrm{[VR_1]}$ and $Y_2=\mathrm{[R_2]}\mathrm{[VR_2]}$ back to $\bar{\bX}_D$ and $\bar{\bX}_F$, then the solution can be rewritten as
 \begin{equation}\label{eqn:yi}
 y_i=a_{i1}\mathrm{[R_1]}+a_{i2}\mathrm{[R_2]}+a_{i3}\mathrm{[R_1]}^2+a_{i4}\mathrm{[R_2]}^2+a_{i5}\mathrm{[R_1]}\mathrm{[VR_1]},
 \end{equation}
 where $i=1,2,\cdots, 11$. We carefully select two solutions
\begin{equation}
\mathrm{[VR_1]}=a_{31}\mathrm{[R_1]}+a_{32}\mathrm{[R_2]}+a_{33}\mathrm{[R_1]}^2+a_{34}\mathrm{[R_2]}^2+a_{35}\mathrm{[R_1]}\mathrm{[VR_1]}
\label{eqn:VR1}
\end{equation}
\begin{equation}
\mathrm{[VR_2]}=a_{41}\mathrm{[R_1]}+a_{42}\mathrm{[R_2]}+a_{43}\mathrm{[R_1]}^2+a_{44}\mathrm{[R_2]}^2+a_{45}\mathrm{[R_1]}\mathrm{[VR_1]}.
\label{eqn:VR2}
\end{equation}
It is easy to solve (\ref{eqn:VR1}) for $\mathrm{[VR_1]}$. Denote the solution as $\mathrm{[VR_1]}=\varphi_1(\mathrm{[R_1]}, \mathrm{[R_2]})$. We then substitute the solution of $\mathrm{[VR_1]}$ to (\ref{eqn:VR2}) and let $\mathrm{[VR_2]}=\varphi_2(\mathrm{[R_1]}, \mathrm{[R_2]})$.
As
\begin{equation}
y_2=\mathrm{[R_2]}\mathrm{[VR_2]}=a_{11,1}\mathrm{[R_1]}+a_{11,2}\mathrm{[R_2]}+a_{11,3}\mathrm{[R_1]}^2+a_{11,4}\mathrm{[R_2]}^2+a_{11,5}\mathrm{[R_1]}\mathrm{[VR_1]},\label{eqn:Y2}
\end{equation}
 we substitute $\mathrm{[VR_1]}=\varphi_1(\mathrm{[R_1]}, \mathrm{[R_2]})$ and $\mathrm{[VR_2]}=\varphi_2(\mathrm{[R_1]}, \mathrm{[R_2]})$ into (\ref{eqn:Y2}), and this will reduce the variables of (\ref{eqn:Y2}) to $\mathrm{[R_1]}$ and $\mathrm{[R_2]}$. The resulting identity is a cubic function that only has $\mathrm{[R_1]}$ and $\mathrm{[R_2]}$ as variables. By Cardano's method, we solve the cubic function symbolically for $\mathrm{[R_2]}$, and denote the only positive real root by $\mathrm{[R_2]}=\psi(\mathrm{[R_1]})$. Substituting $\mathrm{[R_2]}=\psi(\mathrm{[R_1]})$ to $\mathrm{[VR_1]}$ solution, we then have
\[\mathrm{[VR_1]}=\varphi_1(\mathrm{[R_1]}, \mathrm{[R_2]})=\varphi_1(\mathrm{[R_1]}, \psi(\mathrm{[R_1]})).\]
As $\mathrm{[R_2]}$ and $\mathrm{[VR_1]}$ are expressed as algebraic functions with variable $\mathrm{[R_1]}$, all the variables in $X_F=\{\mathrm{[R_1]},\mathrm{[R_2]},\mathrm{[R_1]}^2,\mathrm{[R_2]}^2,\mathrm{[R_1]}\mathrm{[VR_1]}\}$ can be represented as an explicit function of $\mathrm{[R_1]}$, and consequently all the solutions of (\ref{eqn:yi}) can be rewritten as functions of $\mathrm{[R_1]}$.
We expressed all 12 variables as functions of $\mathrm{[R_1]}$. With the conservation law
\begin{eqnarray}
R_{total}=\mathrm{[R_1]+[R_2]+2[RR_1]+2[RR_2]+[VR_1]+[VR_2]+
2[VRR_1]+2[VRR_2]  } \nonumber\\ \mathrm{
+2[RVR_1] +2[RVR_2]+2[\Delta_1]+2[\Delta_2]}~~,
\end{eqnarray}
 we can solve the equation for $\mathrm{[R_1]}$ numerically for any given value of $R_\mathrm{total}$. For all parameter values used in this paper, the dependence on $\Rmono$ was consistent with a single real root,
 leading us to the conclusion that the system had a unique steady state. This does not exclude the possibility of multiple steady states for other parameter sets.

\section{Results and Discussion}

We obtained steady states for the differential equations (\ref{ODE}) by numerically solving the steady state equations as outlined above, for various values of the relative size of the high density area ($f = 5\% \ldots 30 \%$), the attractiveness parameter ($\alpha = 1 \ldots 10$), and for VEGF concentrations ranging from $0.01$ nM to $5$ nM. We considered three situations for the relative mobility of dimers 
$\beta=0.5,0.25,0$ (note that the $\beta = 0.5$ case corresponds to equivalent monomer and dimer mobilities).

\begin{figure}[ht!]
\captionsetup{font=scriptsize}
\begin{center}
\includegraphics[trim=1.2in 0.9in 1.3in 0.9in, clip, width=0.9\textwidth]{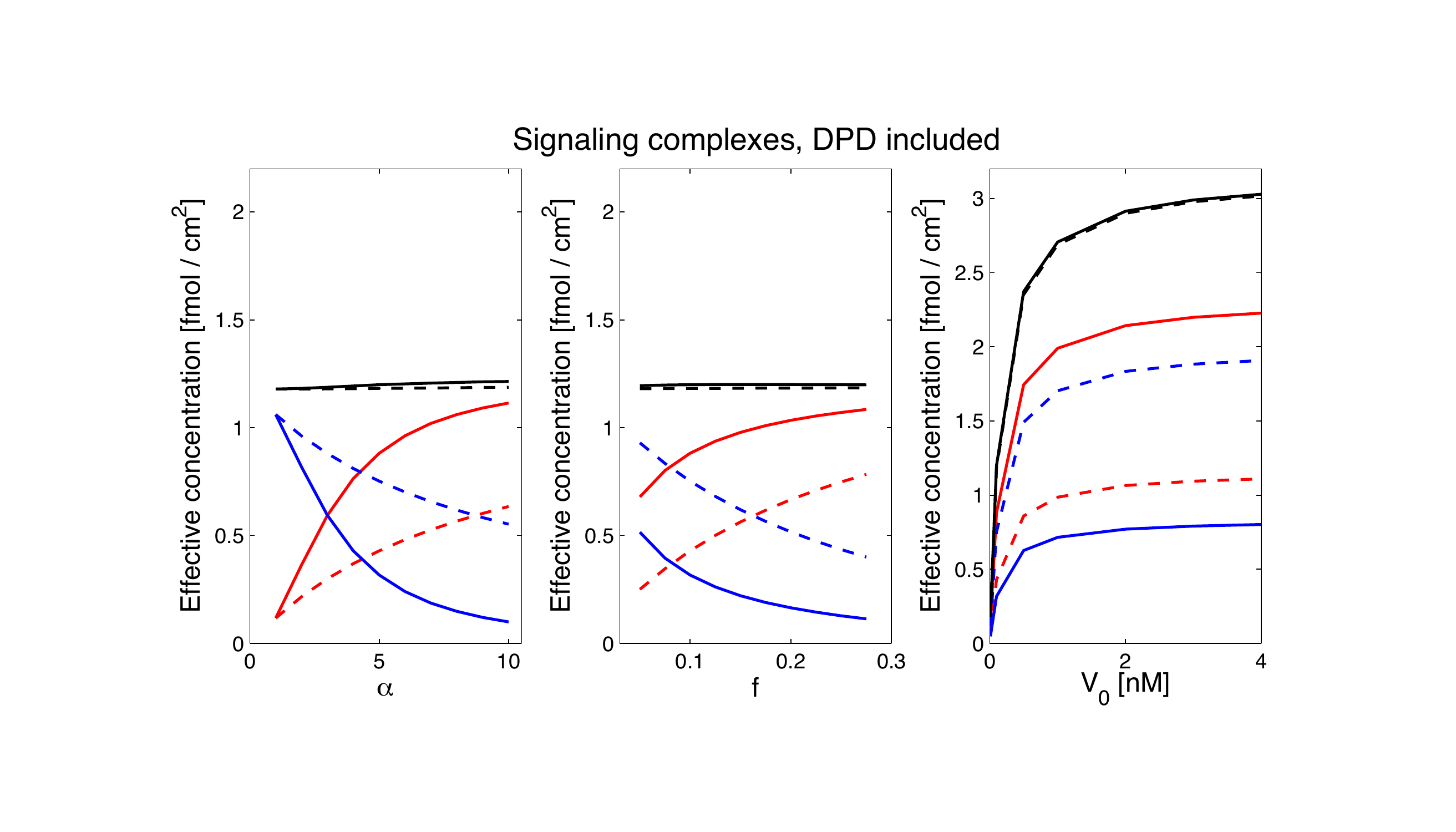}
\end{center}
\vspace{-0.2in}
\caption{Signaling complexes ($RVR$ and $\Delta$) in the high-affinity (red) and normal (blue) domains, as well as total signal (black) as a function of the attractiveness factor $\alpha$, the relative size of the HD domain $f$, and the concentration of VEGF ligand $V_0$. The values for the fixed parameters were $\alpha=5, f=0.1, V_0=0.1$ nM. Solid lines correspond to the case when dimers are not allowed to cross domain boundaries, and dashed lines correspond to fully mobile dimers.
The total signal (black lines) depends weakly on the affinity and size of the attractive domain due to the combined effects of the relatively high ($V_0=0.1$ nM) VEGF concentration value used in the calculations, as well as due to the presence of strong ligand-independent dimerization  (DPD) in the model.
%
}
\label{fig:FigA}
\end{figure}

\begin{figure}[h!]
\captionsetup{font=scriptsize}
\begin{center}
\includegraphics[trim=1.2in 0.9in 1.3in 0.9in, clip, width=0.9\textwidth]{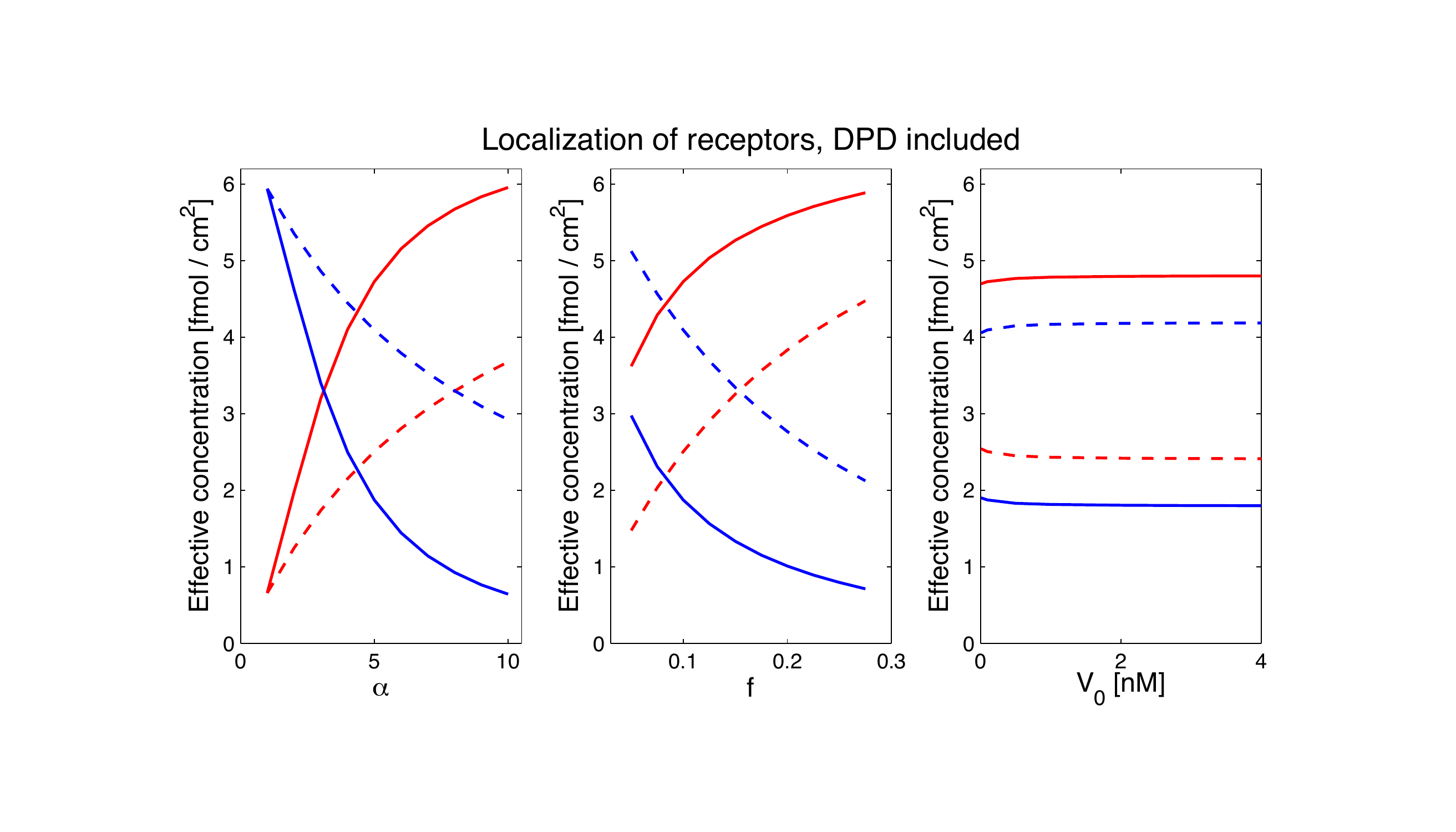}
\end{center}\vspace{-0.2in}
\caption{
Total number of receptors in the high-affinity (red) and normal (blue) domains, in the same set of calculations as in Figure \ref{fig:FigA}.
%
The affinity and size of the HD domain strongly influence clustering (represented by the accumulation of receptors in the HD domain), however, the effect of VEGF is marginal, because the model includes {\em ligand-independent} dimerization (DPD).
}
\label{fig:FigC}
\end{figure}

We first performed calculations using the full model of \cite{MacGabhannP2007}. The equilibrium values for the total number of receptors and signaling complexes in the two domains, as a function of the three parameters ($\alpha$, $f$, $V_0$), are shown in Figures \ref{fig:FigA} and \ref{fig:FigC}. Not surprisingly, increasing the attractiveness parameter $\alpha$, relative size $f$ of the HD area  results in an increasing fraction of receptors and signaling complexes in the high density area. Increasing the concentration of VEGF leads to overall increased singaling but no significant shifts between the domains.

In Figure \ref{fig:FigA}, the total amount of signaling complexes increases only weakly as a function of the attractiveness parameter $\alpha$. This set calculations was performed including both the ligand-induced dimerization as well as the pre-dimerization (DPD) mechanism of the Mac Gabhann-Popel model. The DPD rate constant $b=0.1$ results in a high degree of dimerization (more than 90\% dimers), even in the absence of ligand or a high affinity domain.
%


We were especially interested in the effect of dimerization on the preferential accumulation of receptors. While the results in Fig. \ref{fig:FigA} and \ref{fig:FigC} indicate that the accumulation effect is stronger when dimers are not allowed to cross domain boundaries, the ligand dose response curves (rightmost panels) show only a marginal effect due to the presence of ligand.
\begin{figure}[h!]
\captionsetup{font=scriptsize}
\begin{center}
\includegraphics[trim=1.2in 0.9in 1.3in 0.9in, clip, width=0.9\textwidth]{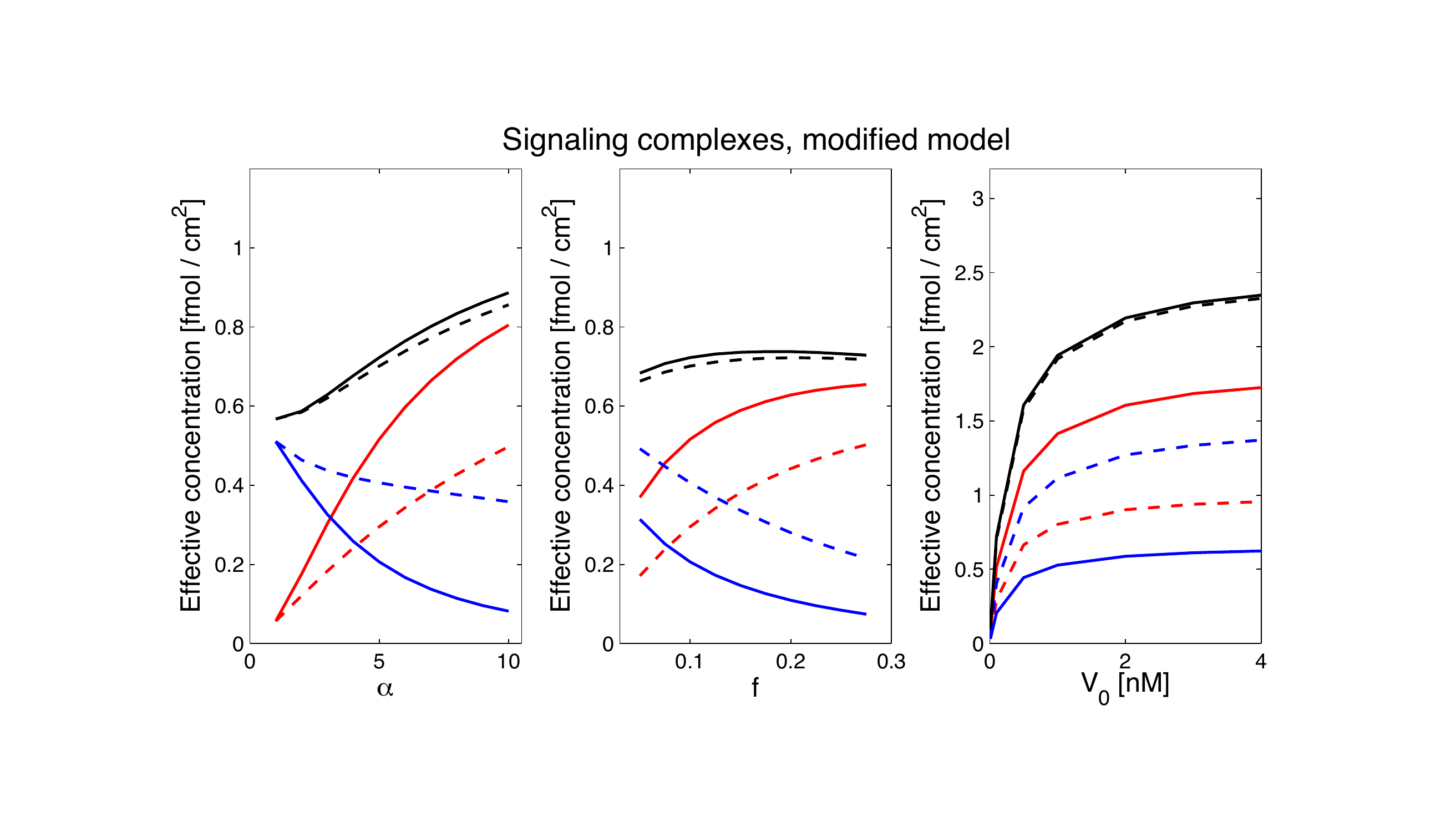}
\end{center}\vspace{-0.2in}
\caption{
Signaling complexes ($RVR$ and $\Delta$) in the high-affinity (red) and normal (blue) domains, as well as total signal (black) as a function of the attractiveness factor $\alpha$, the relative size of the HD domain $f$, and the concentration of VEGF ligand $V_0$. The values for the fixed parameters were $\alpha=5, f=0.1, V_0=0.1$ nM. Solid lines correspond to the case when dimers are not allowed to cross domain boundaries, and dashed lines correspond to fully mobile dimers. This calculation used significantly reduced {\em on-surface dimeriztion rates}, namely $a_s=0.0021$ and $b=0.0001$ (essentially eliminating DPD).
By contrast with Figure \ref{fig:FigA}, the affinity of the HD domain strongly enhances the signal, as dimers are formed at a higher rate in the HD domain.
}
\label{fig:FigD}
\end{figure}

\begin{figure}[h!]
\captionsetup{font=scriptsize}
\begin{center}
\includegraphics[trim=1.2in 0.9in 1.3in 0.9in, clip, width=0.9\textwidth]{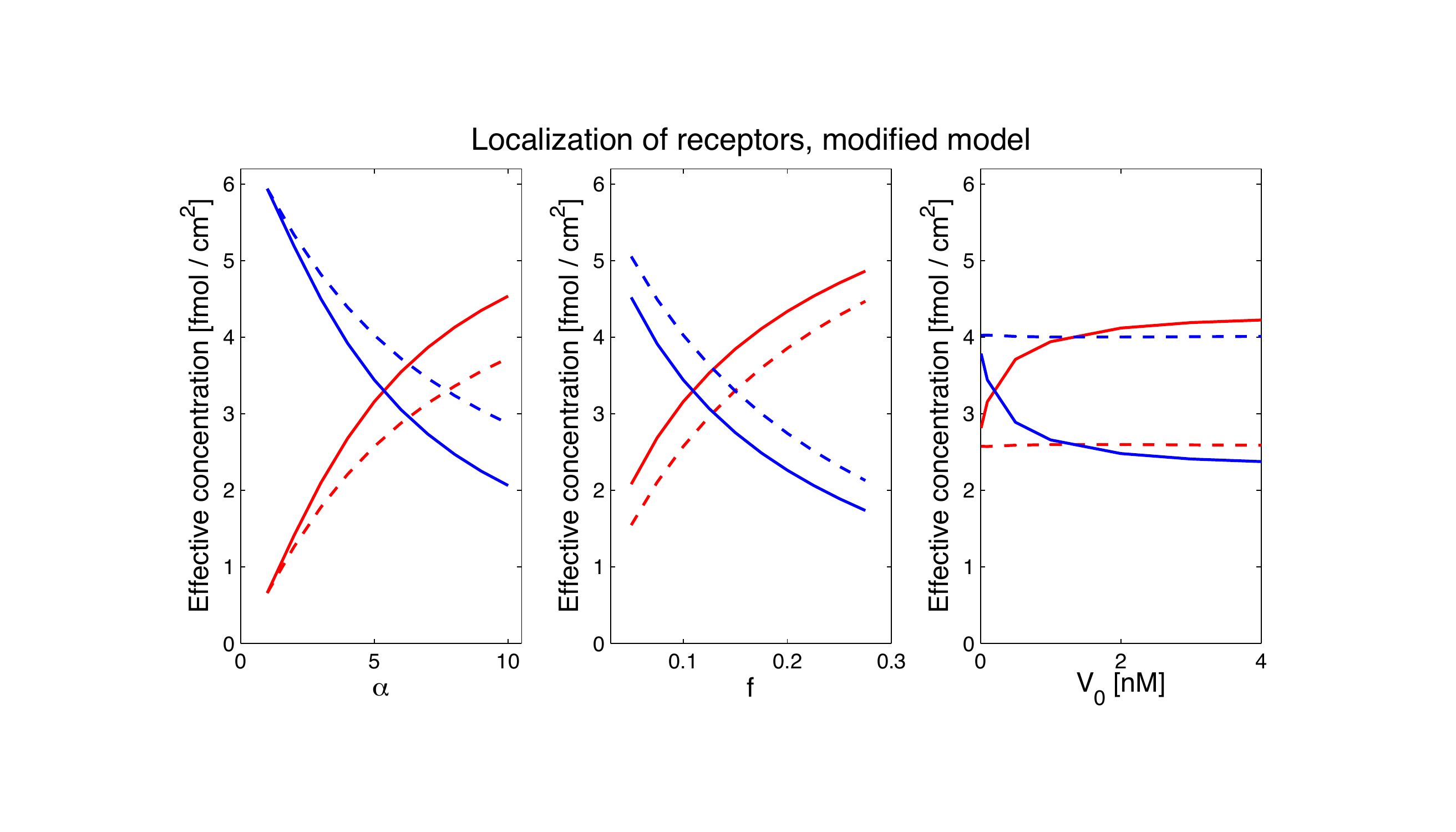}
\end{center}\vspace{-0.2in}
\caption{
Total number of receptors in the high-affinity (red) and normal (blue) domains, in the same set of calculations as in Figure \ref{fig:FigD}.
%
By contrast with Figure \ref{fig:FigC}, the presence of VEGF strongly promotes the accumulation of receptors in the HD domain. Dimers are formed at a higher rate in the HD domain, and become trapped, causing the HD domain to act as a receptor sink.
}
\label{fig:FigF}
\end{figure}

The explanation for the weakness of these effects is the presence of ligand-independent dimerization (or pre-dimerization) in the Mac Gabhann-Popel model, as well as the high value of the on-surface VEGF-receptor binding constant $a_S$. The effect of dimerization on clustering is revealed when pre-dimerization is turned off (by setting $b=0.0001$ in the rate laws) and the on-surface ligand binding rate $a_s$ is reduced.
These results are shown in Fig. \ref{fig:FigD} and \ref{fig:FigF}. The signal is clearly increased as the affinity of the HD domain increases. There is an optimum in the size of the HD area (middle panel, Fig. \ref{fig:FigD}). The effect on the signal persists when dimers are allowed to move.
While the dependence of localization on attractiveness and domain size are similar to the previous set, the depedendence on $V_0$ is dramatically different in Fig. \ref{fig:FigF}, showing a significant increase in the number of receptors in the HD area as $V_0$ is increased. This effect is completely absent when dimers are allowed to move at the same rate as monomers and is very significantly weakened at the intermediate mobility value we used (those results not shown).

In summary, our simple model shows a positive feedback between dimerization / signaling and receptor clustering. Accumulation of receptors in a high affinity patch enhances dimerization and signaling. On the other hand, increased dimerization, in the presence of ligand, inceased the accumulatin of receptors in the high affinity patch. The latter effect also requires a dramatic reduction in the mobility of dimers. Finally, we should point out that the empirically obtained model parameters lead to dimerization rates that are so high that the effects we described here would be marginal; however, the empirically determined parameters are what one would observe {\em as a result of } affinity-induced clustering. In other words, in the presence of high density domains, significanly lower dimerization rate constants may be sufficient to achieve the observed signaling.

\bibliographystyle{eptcs}
\bibliography{generic}

\begin{thebibliography}{10}
\providecommand{\bibitemdeclare}[2]{}
\providecommand{\surnamestart}{}
\providecommand{\surnameend}{}
\providecommand{\urlprefix}{Available at }
\providecommand{\url}[1]{\texttt{#1}}
\providecommand{\href}[2]{\texttt{#2}}
\providecommand{\urlalt}[2]{\href{#1}{#2}}
\providecommand{\doi}[1]{doi:\urlalt{http://dx.doi.org/#1}{#1}}
\providecommand{\bibinfo}[2]{#2}

\bibitemdeclare{article}{AndrewsLPBW2008}
\bibitem{AndrewsLPBW2008}
\bibinfo{author}{N.~L. \surnamestart Andrews\surnameend},
  \bibinfo{author}{K.~A. \surnamestart Lidke\surnameend},
  \bibinfo{author}{J.~R. \surnamestart Pfeiffer\surnameend},
  \bibinfo{author}{A.~R. \surnamestart Burns\surnameend} \&
  \bibinfo{author}{B.~S. \surnamestart Wilson\surnameend}
  (\bibinfo{year}{2008}): \emph{\bibinfo{title}{Actin restricts FCeRI diffusion
  and facilitates antigen-induced receptor immobilization}}.
\newblock {\sl \bibinfo{journal}{Nat Cell Biol}} \bibinfo{volume}{10}, pp.
  \bibinfo{pages}{955--963}, \doi{10.1038/ncb1755}.

\bibitemdeclare{article}{BirkBMC2010}
\bibitem{BirkBMC2010}
\bibinfo{author}{D.~A. \surnamestart Birk\surnameend},
  \bibinfo{author}{J.~\surnamestart Barbato\surnameend},
  \bibinfo{author}{L.~\surnamestart Mureebe\surnameend} \&
  \bibinfo{author}{R.~A. \surnamestart Chaer\surnameend}
  (\bibinfo{year}{2010}): \emph{\bibinfo{title}{Current insights on the biology
  and clinical aspects of VEGF regulation}}.
\newblock {\sl \bibinfo{journal}{Vasc Endovascular Surg}} \bibinfo{volume}{42},
  pp. \bibinfo{pages}{517--530}, \doi{10.1177/1538574408322755}.

\bibitemdeclare{article}{CostaRWVE2009}
\bibitem{CostaRWVE2009}
\bibinfo{author}{M.~N. \surnamestart Costa\surnameend},
  \bibinfo{author}{K.~\surnamestart Radhakrishnan\surnameend},
  \bibinfo{author}{B.~S. \surnamestart Wilson\surnameend},
  \bibinfo{author}{D.~G. \surnamestart Vlachos\surnameend} \&
  \bibinfo{author}{J.~S. \surnamestart Edwards\surnameend}
  (\bibinfo{year}{2009}): \emph{\bibinfo{title}{Coupled Stochastic Spatial and
  Non-Spatial Simulations of ErbB1 Signaling Pathways Demonstrate the
  Importance of Spatial Organization on Signal Transduction}}.
\newblock {\sl \bibinfo{journal}{PLOS One}}
  \bibinfo{volume}{4}(\bibinfo{number}{7}), p. \bibinfo{pages}{6316},
  \doi{10.1371/journal.pone.0006316.s001}.

\bibitemdeclare{article}{FerraraHGN2004}
\bibitem{FerraraHGN2004}
\bibinfo{author}{N.~\surnamestart Ferrara\surnameend}, \bibinfo{author}{K.~J.
  \surnamestart Hilla\surnameend}, \bibinfo{author}{H.~P. \surnamestart
  Gerber\surnameend} \& \bibinfo{author}{W.~\surnamestart Novotny\surnameend}
  (\bibinfo{year}{2004}): \emph{\bibinfo{title}{Discovery anddevelopment of
  bevazicumab, an anti-VEGF antibody for treating cancer}}.
\newblock {\sl \bibinfo{journal}{Net Rev Drug Discov}} \bibinfo{volume}{3}, pp.
  \bibinfo{pages}{391--400}, \doi{10.1038/nrd1381}.

\bibitemdeclare{article}{MacGabhannP2007}
\bibitem{MacGabhannP2007}
\bibinfo{author}{F.~Mac \surnamestart Gabhann\surnameend} \&
  \bibinfo{author}{A.~S. \surnamestart Popel\surnameend}
  (\bibinfo{year}{2007}): \emph{\bibinfo{title}{Dimerization of VEGF receptors
  and implications for signal transduction: A computational study}}.
\newblock {\sl \bibinfo{journal}{Biophysical Chemistry}} \bibinfo{volume}{128},
  pp. \bibinfo{pages}{125--139}, \doi{10.1016/j.bpc.2007.03.010}.

\bibitemdeclare{article}{GallegosSKSB2006}
\bibitem{GallegosSKSB2006}
\bibinfo{author}{A.~M. \surnamestart Gallegos\surnameend},
  \bibinfo{author}{S.~M. \surnamestart Storey\surnameend},
  \bibinfo{author}{A.~B. \surnamestart Kier\surnameend},
  \bibinfo{author}{F.~\surnamestart Shroeder\surnameend} \&
  \bibinfo{author}{J.~M. \surnamestart Ball\surnameend} (\bibinfo{year}{2006}):
  \emph{\bibinfo{title}{Structure and cholesterol dynamics of caveloae/raft and
  nonraft plasma membrane domains}}.
\newblock {\sl \bibinfo{journal}{Biochemistry}} \bibinfo{volume}{45}, pp.
  \bibinfo{pages}{12100--12116}, \doi{10.1021/bi0602720}.

\bibitemdeclare{article}{HalaszLMRE2013}
\bibitem{HalaszLMRE2013}
\bibinfo{author}{\'A.~M. \surnamestart Hal\'asz\surnameend},
  \bibinfo{author}{H.-J. \surnamestart Lai\surnameend},
  \bibinfo{author}{M.~M.~McCabe \surnamestart Pryor\surnameend},
  \bibinfo{author}{K.~\surnamestart Radhakrishnan\surnameend} \&
  \bibinfo{author}{J.~S. \surnamestart Edwards\surnameend}
  (\bibinfo{year}{2013}): \emph{\bibinfo{title}{Analytical Solution of Steady
  State Equations for Chemical Reaction Networks with Bilinear Rate Laws}}.
\newblock {\sl \bibinfo{journal}{IEEE/ACM Transactions on Computational Biology
  and Bioinformatics}}, \doi{10.1109/TCBB.2013.41}.
\newblock \bibinfo{note}{Epub before print}.

\bibitemdeclare{article}{HanahanF1996}
\bibitem{HanahanF1996}
\bibinfo{author}{D.~\surnamestart Hanahan\surnameend} \&
  \bibinfo{author}{J.~\surnamestart Folkman\surnameend} (\bibinfo{year}{1996}):
  \emph{\bibinfo{title}{Patterns and emerging mechanisms of the angiogenetic
  switch during tumorigenesis}}.
\newblock {\sl \bibinfo{journal}{Cell}} \bibinfo{volume}{86}, pp.
  \bibinfo{pages}{353--364}, \doi{10.1016/S0092-8674(00)80108-7}.

\bibitemdeclare{article}{HsiehYRSVSWE2008}
\bibitem{HsiehYRSVSWE2008}
\bibinfo{author}{M.-Y. \surnamestart Hsieh\surnameend},
  \bibinfo{author}{S.~\surnamestart Yang\surnameend},
  \bibinfo{author}{M.~\surnamestart Raymond-Stinz\surnameend},
  \bibinfo{author}{S.~\surnamestart Steinberg\surnameend},
  \bibinfo{author}{D.~\surnamestart Vlachos\surnameend},
  \bibinfo{author}{W.~\surnamestart Shu\surnameend},
  \bibinfo{author}{B.~\surnamestart Wilson\surnameend} \&
  \bibinfo{author}{J.~S. \surnamestart Edwards\surnameend}
  (\bibinfo{year}{2008}): \emph{\bibinfo{title}{Stochastic Simulations of ErbB
  Homo- and Hetero-Dimerization: Potential IMpacts of Receptor Conformational
  State and Spatial Segregation}}.
\newblock {\sl \bibinfo{journal}{IET Systems Biology}} \bibinfo{volume}{2}, pp.
  \bibinfo{pages}{256--272}, \doi{10.1049/iet-syb:20070073}.

\bibitemdeclare{article}{Karamysheva2008}
\bibitem{Karamysheva2008}
\bibinfo{author}{A.~F. \surnamestart Karamysheva\surnameend}
  (\bibinfo{year}{2008}): \emph{\bibinfo{title}{Mechanisms of angiogenesis}}.
\newblock {\sl \bibinfo{journal}{Biochemistry (Mosc)}} \bibinfo{volume}{73},
  pp. \bibinfo{pages}{751--762}, \doi{10.1134/S0006297908070031}.

\bibitemdeclare{article}{KleimanMCLS2011}
\bibitem{KleimanMCLS2011}
\bibinfo{author}{L.~B. \surnamestart Kleiman\surnameend},
  \bibinfo{author}{T.~\surnamestart Maiwald\surnameend},
  \bibinfo{author}{H.~\surnamestart Conzelman\surnameend},
  \bibinfo{author}{D.~A. \surnamestart Lauffenburger\surnameend} \&
  \bibinfo{author}{P.~K. \surnamestart Sorger\surnameend}
  (\bibinfo{year}{2011}): \emph{\bibinfo{title}{Rapid phospho-turnover by
  receptor tyrosine kinases impacts downstream signaling and drug binding}}.
\newblock {\sl \bibinfo{journal}{Molecular Cell}} \bibinfo{volume}{43}, p.
  \bibinfo{pages}{723¨C737}, \doi{10.1016/j.molcel.2011.07.014}.

\bibitemdeclare{article}{LemonS2010}
\bibitem{LemonS2010}
\bibinfo{author}{M.~A. \surnamestart Lemon\surnameend} \&
  \bibinfo{author}{J.~\surnamestart Schlessinger\surnameend}
  (\bibinfo{year}{2010}): \emph{\bibinfo{title}{Cell Signaling by Receptor
  Tyrosine Kinases}}.
\newblock {\sl \bibinfo{journal}{Cell}} \bibinfo{volume}{141}, pp.
  \bibinfo{pages}{1117--1134}, \doi{10.1016/j.cell.2010.06.011}.

\bibitemdeclare{article}{LillemeierPSWD2006}
\bibitem{LillemeierPSWD2006}
\bibinfo{author}{B.~F. \surnamestart Lillemeier\surnameend},
  \bibinfo{author}{J.~R. \surnamestart Pfeiffer\surnameend},
  \bibinfo{author}{Z.~\surnamestart Surviladze\surnameend},
  \bibinfo{author}{B.~S. \surnamestart Wilson\surnameend} \&
  \bibinfo{author}{M.~M. \surnamestart Davis\surnameend}
  (\bibinfo{year}{2006}): \emph{\bibinfo{title}{Plasma membrane-associated
  proteins are clustered into islands attached to the cytoskeleton}}.
\newblock {\sl \bibinfo{journal}{Proc Natl Acad Sci USA}}
  \bibinfo{volume}{103}, pp. \bibinfo{pages}{18992--18997},
  \doi{10.1073/pnas.0609009103}.

\bibitemdeclare{article}{LohelaBTA2009}
\bibitem{LohelaBTA2009}
\bibinfo{author}{M.~\surnamestart Lohela\surnameend},
  \bibinfo{author}{M.~\surnamestart Bry\surnameend},
  \bibinfo{author}{T.~\surnamestart Tammela\surnameend} \&
  \bibinfo{author}{K.~\surnamestart Alitalo\surnameend} (\bibinfo{year}{2009}):
  \emph{\bibinfo{title}{VEGFs and receptors involved in angiogenesis versus
  lymphangiogenesis}}.
\newblock {\sl \bibinfo{journal}{Curr Opin Cell Biol}} \bibinfo{volume}{21},
  pp. \bibinfo{pages}{154--165}, \doi{10.1016/j.ceb.2008.12.012}.

\bibitemdeclare{article}{MayawalaVE22005}
\bibitem{MayawalaVE22005}
\bibinfo{author}{K.~\surnamestart Mayawala\surnameend}, \bibinfo{author}{D.~G.
  \surnamestart Vlachos\surnameend} \& \bibinfo{author}{J.~S. \surnamestart
  Edwards\surnameend} (\bibinfo{year}{2005}):
  \emph{\bibinfo{title}{Computational modeling reveals molecular details of
  epidermal growth factor binding}}.
\newblock {\sl \bibinfo{journal}{BMC Cell Biol}} \bibinfo{volume}{6},
  p.~\bibinfo{pages}{41}, \doi{10.1186/1471-2121-6-41}.

\bibitemdeclare{article}{MayawalaVE2005}
\bibitem{MayawalaVE2005}
\bibinfo{author}{K.~\surnamestart Mayawala\surnameend}, \bibinfo{author}{D.~G.
  \surnamestart Vlachos\surnameend} \& \bibinfo{author}{J.~S. \surnamestart
  Edwards\surnameend} (\bibinfo{year}{2005}):
  \emph{\bibinfo{title}{Heterogeneities in EGF rceptor density at the cell
  surface can lead to concave up Scatchard plot of EGF binding}}.
\newblock {\sl \bibinfo{journal}{FEBS Letters}} \bibinfo{volume}{579}, pp.
  \bibinfo{pages}{3043--3047}, \doi{10.1016/j.febslet.2005.04.059}.

\bibitemdeclare{article}{MingYREW2010}
\bibitem{MingYREW2010}
\bibinfo{author}{H.~\surnamestart Ming-Yu\surnameend},
  \bibinfo{author}{S.~\surnamestart Yang\surnameend}, \bibinfo{author}{M.~A.
  \surnamestart Raymond-Stinz\surnameend}, \bibinfo{author}{J.~S. \surnamestart
  Edwards\surnameend} \& \bibinfo{author}{B.~S. \surnamestart
  Wilson\surnameend} (\bibinfo{year}{2010}):
  \emph{\bibinfo{title}{Spatiotemporal modeling of signalingprotein recruitment
  to EGFR}}.
\newblock {\sl \bibinfo{journal}{BMC Systems Biology}} \bibinfo{volume}{4},
  p.~\bibinfo{pages}{57}.

\bibitemdeclare{article}{MuraseFUSIM2004}
\bibitem{MuraseFUSIM2004}
\bibinfo{author}{K.~\surnamestart Murase\surnameend},
  \bibinfo{author}{T.~\surnamestart Fujiwara\surnameend},
  \bibinfo{author}{Y.~\surnamestart Umemura\surnameend},
  \bibinfo{author}{K.~\surnamestart Suzuki\surnameend},
  \bibinfo{author}{R.~\surnamestart Iino\surnameend} \&
  \bibinfo{author}{H.~\surnamestart Murakoshi\surnameend}
  (\bibinfo{year}{2004}): \emph{\bibinfo{title}{Ultrafine membrane compartments
  for molecular diffusion as revealed by single molecule techniques}}.
\newblock {\sl \bibinfo{journal}{Biophys J}} \bibinfo{volume}{75}, pp.
  \bibinfo{pages}{4075--4093}, \doi{10.1529/biophysj.103.035717}.

\bibitemdeclare{article}{OlssonKC2006}
\bibitem{OlssonKC2006}
\bibinfo{author}{D.~A. \surnamestart Olsson\surnameend},
  \bibinfo{author}{J.~\surnamestart Kreuger\surnameend} \&
  \bibinfo{author}{L.~\surnamestart Claesson-Welsh\surnameend}
  (\bibinfo{year}{2006}): \emph{\bibinfo{title}{VEGF receptor signaling - in
  control of vascular function}}.
\newblock {\sl \bibinfo{journal}{Nat Rev Mol Cell Biol}} \bibinfo{volume}{7},
  pp. \bibinfo{pages}{359--371}, \doi{10.1038/nrm1911}.

\bibitemdeclare{article}{ORRHOOW2005}
\bibitem{ORRHOOW2005}
\bibinfo{author}{G.~\surnamestart Orr\surnameend},
  \bibinfo{author}{D.~\surnamestart Hu\surnameend},
  \bibinfo{author}{S.~\surnamestart Ozcelik\surnameend}, \bibinfo{author}{L.~K.
  \surnamestart Opresko\surnameend} \& \bibinfo{author}{H.~S. \surnamestart
  Wiley\surnameend} (\bibinfo{year}{2005}): \emph{\bibinfo{title}{Cholesterol
  dictates the freedom of EGF receptors and HER2 in the plane of the
  membrane}}.
\newblock {\sl \bibinfo{journal}{Biophys J}} \bibinfo{volume}{89}, pp.
  \bibinfo{pages}{1362--1373}, \doi{10.1529/biophysj.104.056192}.

\bibitemdeclare{article}{PlouetSG1989}
\bibitem{PlouetSG1989}
\bibinfo{author}{J.~\surnamestart Plouet\surnameend},
  \bibinfo{author}{J.~\surnamestart Schilling\surnameend} \&
  \bibinfo{author}{D.~\surnamestart Gospodarowicz\surnameend}
  (\bibinfo{year}{1989}): \emph{\bibinfo{title}{Isolation and characterization
  of a newly identified endothelial cell mitogen produced by AtT-20 cels}}.
\newblock {\sl \bibinfo{journal}{EMBO J}} \bibinfo{volume}{8}, pp.
  \bibinfo{pages}{3801--3806}.

\bibitemdeclare{article}{RitchieK2003}
\bibitem{RitchieK2003}
\bibinfo{author}{K.~\surnamestart Ritchie\surnameend} \&
  \bibinfo{author}{A.~\surnamestart Kusumi\surnameend} (\bibinfo{year}{2003}):
  \emph{\bibinfo{title}{Single-particle tracking image microscopy}}.
\newblock {\sl \bibinfo{journal}{Methods Enzymol}} \bibinfo{volume}{360}, pp.
  \bibinfo{pages}{618--634}, \doi{10.1016/S0076-6879(03)60131-X}.

\bibitemdeclare{article}{Roskoski2008}
\bibitem{Roskoski2008}
\bibinfo{author}{R.~J. \surnamestart Roskoski\surnameend}
  (\bibinfo{year}{2008}): \emph{\bibinfo{title}{VEGF receptor protein-tyrosine
  kinases: structure and regulation}}.
\newblock {\sl \bibinfo{journal}{Biochem Biophys Res Commun}}
  \bibinfo{volume}{375}, pp. \bibinfo{pages}{287--291},
  \doi{10.1016/j.bbrc.2008.07.121}.

\bibitemdeclare{article}{SchorederGASM2001}
\bibitem{SchorederGASM2001}
\bibinfo{author}{F.~\surnamestart Schoreder\surnameend}, \bibinfo{author}{A.~M.
  \surnamestart Gallegos\surnameend}, \bibinfo{author}{B.~P. \surnamestart
  Atshaves\surnameend}, \bibinfo{author}{S.~M. \surnamestart Storey\surnameend}
  \& \bibinfo{author}{A.~L. \surnamestart McIntosh\surnameend}
  (\bibinfo{year}{2001}): \emph{\bibinfo{title}{Recent advances in membrane
  microdomains: rafts, caveolae, and intracellular cholesterol trafficking}}.
\newblock {\sl \bibinfo{journal}{Exp Bio9l Med (Maywood)}}
  \bibinfo{volume}{226}, pp. \bibinfo{pages}{873--890}.

\bibitemdeclare{article}{SengerGDPHD1983}
\bibitem{SengerGDPHD1983}
\bibinfo{author}{D.~R. \surnamestart Senger\surnameend}, \bibinfo{author}{S.~J.
  \surnamestart Galli\surnameend}, \bibinfo{author}{A.~M. \surnamestart
  Dvorak\surnameend}, \bibinfo{author}{C.~A. \surnamestart
  Perruzzi\surnameend}, \bibinfo{author}{V.~S. \surnamestart Harvey\surnameend}
  \& \bibinfo{author}{H.~F. \surnamestart Dvorak\surnameend}
  (\bibinfo{year}{1983}): \emph{\bibinfo{title}{Tumor cells secrete a vascular
  permeability factor that promotes accumulation of ascites fluid}}.
\newblock {\sl \bibinfo{journal}{Science}} \bibinfo{volume}{219}, pp.
  \bibinfo{pages}{983--985}, \doi{10.1126/science.6823562}.

\bibitemdeclare{article}{SingerN1972}
\bibitem{SingerN1972}
\bibinfo{author}{S.~J. \surnamestart Singer\surnameend} \&
  \bibinfo{author}{G.~L. \surnamestart Nicholson\surnameend}
  (\bibinfo{year}{1972}): \emph{\bibinfo{title}{The Fluid Mosaic Model of the
  Structure of Cell Membranes}}.
\newblock {\sl \bibinfo{journal}{Science}} \bibinfo{volume}{175}, pp.
  \bibinfo{pages}{720--731}, \doi{10.1126/science.175.4023.720}.

\bibitemdeclare{article}{StuttfildB2009}
\bibitem{StuttfildB2009}
\bibinfo{author}{E.~\surnamestart Stuttfield\surnameend} \&
  \bibinfo{author}{K.~\surnamestart Ballmer-Hofer\surnameend}
  (\bibinfo{year}{2009}): \emph{\bibinfo{title}{Structure and Function of VEGF
  Receptors}}.
\newblock {\sl \bibinfo{journal}{IUBMB Life}}
  \bibinfo{volume}{61}(\bibinfo{number}{9}), pp. \bibinfo{pages}{915--922},
  \doi{10.1002/iub.234}.

\bibitemdeclare{article}{VerbSMNFVMWD2003}
\bibitem{VerbSMNFVMWD2003}
\bibinfo{author}{G.~\surnamestart Vereb\surnameend},
  \bibinfo{author}{J.~\surnamestart Szollosi\surnameend},
  \bibinfo{author}{J.~\surnamestart Matko\surnameend},
  \bibinfo{author}{P.~\surnamestart Nagy\surnameend},
  \bibinfo{author}{T.~\surnamestart Farkas\surnameend},
  \bibinfo{author}{L.~\surnamestart Vigh\surnameend},
  \bibinfo{author}{L.~\surnamestart Matyus\surnameend}, \bibinfo{author}{T.~A.
  \surnamestart Waldmann\surnameend} \& \bibinfo{author}{S.~\surnamestart
  Djanovich\surnameend} (\bibinfo{year}{2003}): \emph{\bibinfo{title}{Dynamic,
  yet structured: The cell membrane threedecades after the Singer-Nicolson
  model}}.
\newblock {\sl \bibinfo{journal}{Proc Natl Acad Sci USA}}
  \bibinfo{volume}{100}(\bibinfo{number}{14}), pp. \bibinfo{pages}{8053--8058},
  \doi{10.1073/pnas.1332550100}.

\bibitemdeclare{article}{WilsonPRLA2007}
\bibitem{WilsonPRLA2007}
\bibinfo{author}{B.~S. \surnamestart Wilson\surnameend}, \bibinfo{author}{J.~R.
  \surnamestart Pfeiffer\surnameend}, \bibinfo{author}{M.~A. \surnamestart
  Raymond-Stintz\surnameend}, \bibinfo{author}{D.~\surnamestart
  Lidke\surnameend} \& \bibinfo{author}{N.~\surnamestart Andrews\surnameend}
  (\bibinfo{year}{2007}): \emph{\bibinfo{title}{Exploring membrane domains
  using native membrane sheets and transmission electron microscopy}}.
\newblock {\sl \bibinfo{journal}{Methods Mol Biol}} \bibinfo{volume}{398}, pp.
  \bibinfo{pages}{245--261}, \doi{10.1007/978-1-59745-513-8\_17}.

\bibitemdeclare{article}{YangRYZLSWOW2007}
\bibitem{YangRYZLSWOW2007}
\bibinfo{author}{S.~\surnamestart Yang\surnameend}, \bibinfo{author}{M.~A.
  \surnamestart Raymond-Stinz\surnameend}, \bibinfo{author}{W.~\surnamestart
  Ying\surnameend}, \bibinfo{author}{J.~\surnamestart Zhang\surnameend},
  \bibinfo{author}{D.~S. \surnamestart Lidke\surnameend},
  \bibinfo{author}{S.~L. \surnamestart Steinberg\surnameend},
  \bibinfo{author}{L.~\surnamestart Williams\surnameend},
  \bibinfo{author}{J.~M. \surnamestart Oliver\surnameend} \&
  \bibinfo{author}{B.~S. \surnamestart Wilson\surnameend}
  (\bibinfo{year}{2007}): \emph{\bibinfo{title}{Mapping ErbB rceptors on breast
  cancer cell membranes during signal transduction}}.
\newblock {\sl \bibinfo{journal}{J Cell Sci}} \bibinfo{volume}{120}, pp.
  \bibinfo{pages}{2763--2773}, \doi{10.1242/jcs.007658}.

\end{thebibliography}
\end{document}